%% file: lineros_phd_ncb.tex
\documentclass[rivista]{cimento} 

\usepackage[utf8]{inputenc}
\usepackage[english]{babel}
\usepackage{color}
\usepackage{ae,aecompl,aeguill} 
\usepackage{amsmath,amssymb,amsfonts}
\usepackage{wasysym}
\usepackage{makeidx,showidx}
\usepackage[toc,page]{appendix}
\usepackage{multirow,rotating}
\usepackage{graphicx}

\input{tex/newcommands.tex}

\graphicspath{{./figs/}{./figs/f1/}{./figs/f2/}{./figs/f3/}{./figs/f4/}{./figs/f5/}}

\title{Positrons from cosmic rays interactions and dark matter annihilations}
\author{Roberto~Alfredo~Lineros~Rodriguez\from{ins:unito}}
\instlist{\inst{ins:unito} Dipartimento di Fisica Teorica, Universit\'{a} degli Studi di Torino, and INFN -- Sezione Torino, via P. Giuria 1, I-10125, Torino, Italy.}
\PACSes{\PACSit{95.35.+d}{96.50.S-}{98.38.-j}{98.70.Sa}}
\begin{document}

\maketitle

\input{tex/abs.tex}

\input{tex/sect1.tex}

\input{tex/sect2.tex}
\input{tex/sect3.tex}

\input{tex/sect4.tex}
\input{tex/sect5.tex}

\input{tex/sect6.tex}

\input{tex/ack.tex}
\input{tex/bib.tex}

\end{document}

%% file: tex/newcommands.tex
\newcommand{\tu}[1]{\textnormal{#1}}             
\newcommand{\tn}[1]{\textnormal{#1}}         
\newcommand{\mb}[1]{\mathbf{#1}}             
\newcommand{\ics}{\langle\eta\sigma\rangle}  

\newcommand{\fluxe}{\Phi_{e^+}}

\newcommand{\mass}[1]{ m_{#1} }
\newcommand{\ener}[1]{ \varepsilon_{#1} }

\newcommand{\dnde}[1]{\frac{dn}{d\ener{#1}}}

\newcommand{\nue}{\nu_{e}}
\newcommand{\nuebar}{\bar{\nu}_{e}}
\newcommand{\numu}{\nu_{\mu}}
\newcommand{\numubar}{\bar{\nu}_{\mu}}
\newcommand{\nutau}{\nu_{\tau}}
\newcommand{\nutaubar}{\bar{\nu}_{\tau}}

\newcommand{\fref}[1]{{\small\textnormal{\ref{#1}}}}
\newcommand{\citeeq}[1]{{\small\textnormal{Equation}}~\fref{#1}}

\newcommand{\citefig}[1]{{\small\textnormal{Figure}}~\fref{#1}}

\newcommand{\citetab}[1]{{\small\textnormal{Table}}~\fref{#1}}

\newcommand{\citesec}[1]{{\small\textnormal{Section}}~\fref{#1}}

\newenvironment{fig}{\begin{figure}[tpb]\vspace{1ex}\begin{minipage}{\textwidth}\centering}{\end{minipage}\end{figure}}
\newenvironment{tab}{\begin{table}[bpt]\vspace{1ex}\begin{minipage}{\textwidth}\centering}{\end{minipage} \end{table}}

\newenvironment{eq}{\begin{eqnarray}}{\end{eqnarray}}

\let\oldsqrt\sqrt
\def\sqrt{\mathpalette\DHLhksqrt}
\def\DHLhksqrt#1#2{%
\setbox0=\hbox{$#1\oldsqrt{#2\,}$}\dimen0=\ht0
\advance\dimen0-0.2\ht0
\setbox2=\hbox{\vrule height\ht0 depth -\dimen0}%
{\box0\lower0.4pt\box2}}

%% file: tex/abs.tex
\begin{abstract}

The electron and positron cosmic rays puzzle has triggered a revolution in the field of astroparticle physics. 
Many hypotheses have been proposed to explain the unexpected rise of the positron fraction, observed by HEAT and PAMELA experiments, for energies larger than few GeVs.  
In this work, we study sources of positron cosmic rays related to annihilation of dark matter and secondary production.
In both cases, we consider the impact of uncertainties related to dark matter physics, nuclear physics and propagation of cosmic rays, finding that the largest uncertainties come from propagation. 
We find that some key--features present in the positron signal from dark matter annihilation are preserved even though the uncertainties. 
In addition, we study the stability of the positron fraction under small variations of the electron flux, which is usually considered as known, we found that considering just the observational uncertainties in the electron flux is enough to changed dramatically the positron fraction in the energy range were the excess is observed.\\

\end{abstract}

%% file: tex/sect1.tex
Preprint {DFTT 2/2010}\\

\section{Introduction}
\label{sec:01}

During the last decades, the astrophysical and cosmological evidences of dark matter and dark energy have created a revolution in the field of fundamental physics.
Some models of new physics, which aim to extend the Standard Model of particles, predict particles with the right properties to act as dark matter candidates.
In a similar way, cosmic ray physics is continuously stimulated with new observations showing an environment much more active than it was expected.
Furthermore, cosmic rays correspond to genuine samples of the matter composition of the Milky Way and give essential information about the Interstellar medium and the Solar System environment.
A fraction of comic rays is composed by antimatter.
This component may provide crucial clues on the non--standard sources of cosmic rays because it is less abundant than the matter cosmic--rays component. 
This analysis may open a window for dark matter searches.\\

This paper is based on results of the author's Ph.D. Thesis \cite{Lineros:2008hh} and some other projects derived from it.
In \citesec{sec:02}, we study the standard mechanisms of production of electrons and positrons. 
In the same spirit, in \citesec{sec:03}, the propagation of cosmic rays is reviewed and discussed.
Then, we study the production of positrons in scenarios of dark matter annihilation (Section \ref{sec:04}) and secondary production (Section \ref{sec:05}).\\

%% file: tex/sect2.tex
\section{Production of positrons and electrons}
\label{sec:02}

In a first approach, the Standard Model of particles physics describes well enough the physics behind the production of cosmic rays. 
However, let us clarify that, the production in the astrophysical context arises both from particles physics and from the galactic environment.\\

Among the sources of cosmic rays, supernovae are the main ones and a big fraction of them comes from the matter expelled at the time of the explosion.
This mechanism is generally the dominant one but not for all the species and energy scales. 
In fact, interactions between cosmic rays and the interstellar gas sizeably contributes, adding an extra amount of cosmic rays known as \emph{secondaries}.\\

Electrons and positrons are also produced by other mechanisms, like pulsar emission~\cite{Busching:2008}, and decay/annihilation of dark matter~\cite{Hooper:2004bq}.\\

\subsection{\sc Production in proton--proton collisions}

The nuclear scattering is the fundamental process behind secondary electrons and positrons. 
The secodary production takes place in the galactic disk when nuclei cosmic rays scatter off the interstellar medium.
In particular, it depends on the composition of the interstellar medium -- composed by hydrogen $(0.9\,\tu{part.}/\tu{cm}^3)$ and helium gas $(0.1\,\tu{part.}/\tu{cm}^3)$~\cite{Thorndike:1930, Field:1969, Lyman:1998,  Ferriere:2001rg} -- and on the fluxes of protons and alpha particles~\cite{Shikaze:2006je}.

%


%
%
%

\subsubsection{\sc Production from mesons decay}

In proton-proton collisions, charged pions and kaons are produced in big amounts.
Their decay channels are responsible of the big amount of electrons and positrons.
Of course, this is valid for collisions at low energy ($p_{\tn{lab}} < 20\,\tu{GeV}$) when meson production is dominant.\\

The electron and positron energy spectra are calculated from the inclusive cross section of pions and kaons and from the electron and positron spectra associated to their decays:
\begin{eq}
 \frac{d\ics_{pp \rightarrow e}}{d\ener{e}}  = \frac{d\ics_{pp \rightarrow e}^{\pi}}{d\ener{e}} + \frac{d\ics_{pp \rightarrow e}^{K}}{d\ener{e}} \; ,
\end{eq}
where a particular contribution is generically calculated as:
\begin{eq}
 \frac{d\ics_{pp \rightarrow e}^{X}}{d\ener{e}}(\ener{e},\ener{p}) = \int d \ener{X} \frac{d \ics_{pp \rightarrow X}}{d\ener{X}}(\ener{X}, \ener{p}) \times f_{e, X}(\ener{e}, \ener{X})  \; ,
\end{eq}
where $X$ denotes pions or kaons as intermediate particles and $f_{e,X}$ the meson decay distribution into electrons or positrons.
The decay distributions is analytically calculated including muon polarization effects~\cite{Lineros:2008hh}. Pion decays are the simplest ones because they have just one dominant decay mode ($\mu^{\pm} \nu_{\mu}$) instead of kaon decays which are complex to treat (details in appendices C and D of~\cite{Lineros:2008hh}). \\

The inclusive cross sections of pions and kaons from proton--proton collisions are very difficult to calculate \emph{a priori}.
In the literature, two well known parametrization for the inclusive cross section of pions and kaons are available: 
\begin{itemize}  
\item Badhwar et al. parametrization~\cite{Badhwar:1977zf,Stephens:1981Ap}.
\item Tan--Ng parametrization~\cite{Tan:1983ICRC,Tan:1984ha}.
\end{itemize}
Both parametrization reproduce good enough the observations in colliders.
In any case, the small differences among them gives an estimation of uncertainties coming from nuclear and particle physics.\\

\subsubsection{\sc Kamae et al. parametrization}

A parametrization of production of positrons, electrons and other particles is proposed by Kamae et al.~\cite{Abe:2004gp,Kamae:2004xx,Kamae:2006bf}. It aims to provide an easy way to compute and estimate cosmic ray fluxes that comes from interstellar medium interactions with nuclei cosmic rays. \\

New processes are included like contributions from $\Delta(1238)$ and many hadron resonances around 1600 $[\tu{MeV}/\tu{c}^2]$ that make it accurate in the very low proton energy range ($p_{\tn{lab}} < 8\,\tu{GeV}$).
Other included processes come from diffractive dissociation which contributes in the intermediate energy range ($p_{\tn{lab}} > 20\,\tu{GeV}$).\\

In addition, this parametrization works for very high energy scales being based on simulators like PYTHIA~\cite{Sjostrand:2000wi} to generate the spectra at higher energies. Also, its big advantage is to save CPU time by avoiding to calculate the convolution among mesons production and meson decay into electrons and positrons.\\

\subsubsection{\sc Uncertainties from nuclear physics}

\begin{fig}
 \resizebox{\hsize}{!}{\includegraphics[angle=270, width=0.5\textwidth]{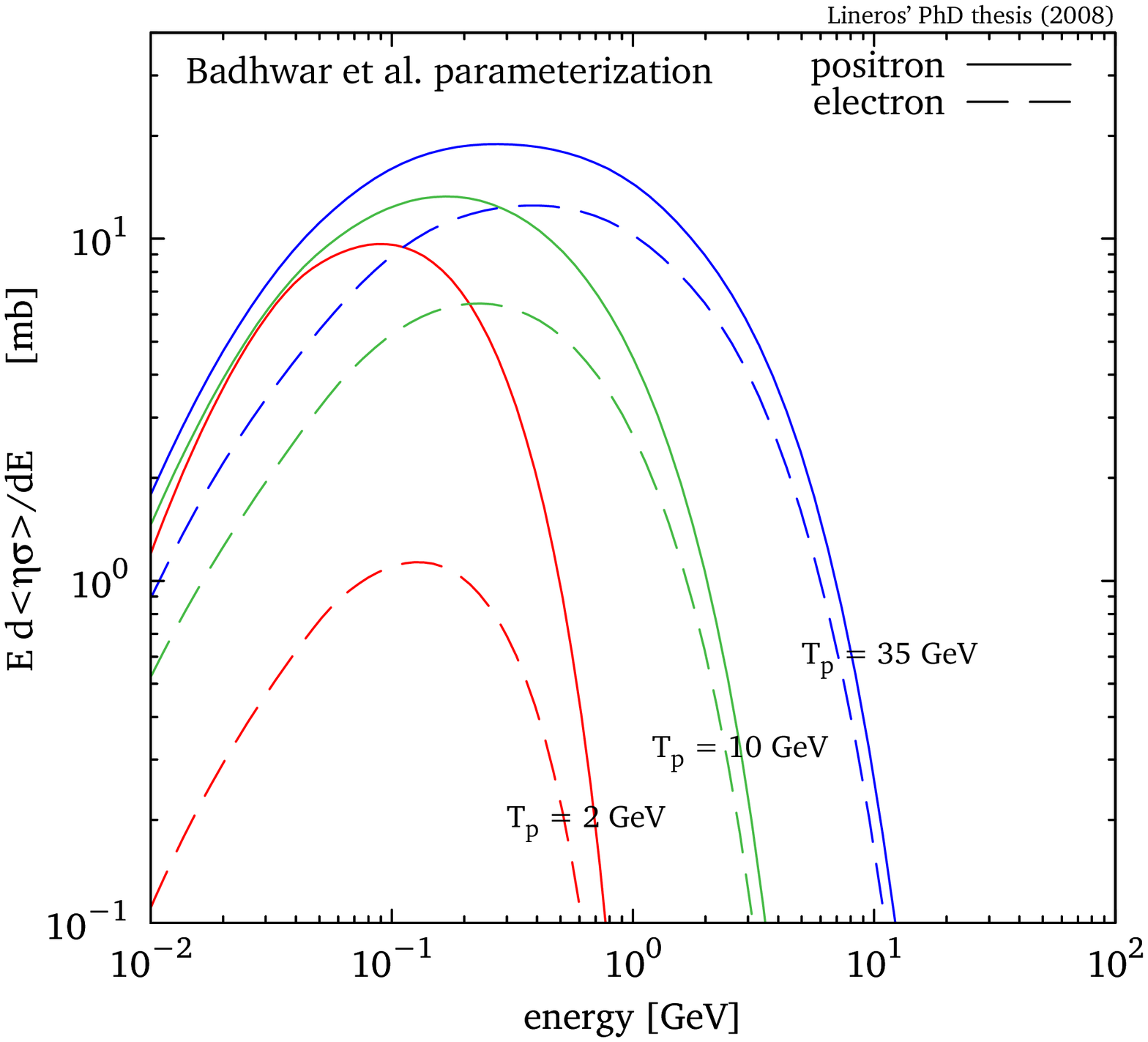} \includegraphics[angle=270, width=0.5\textwidth]{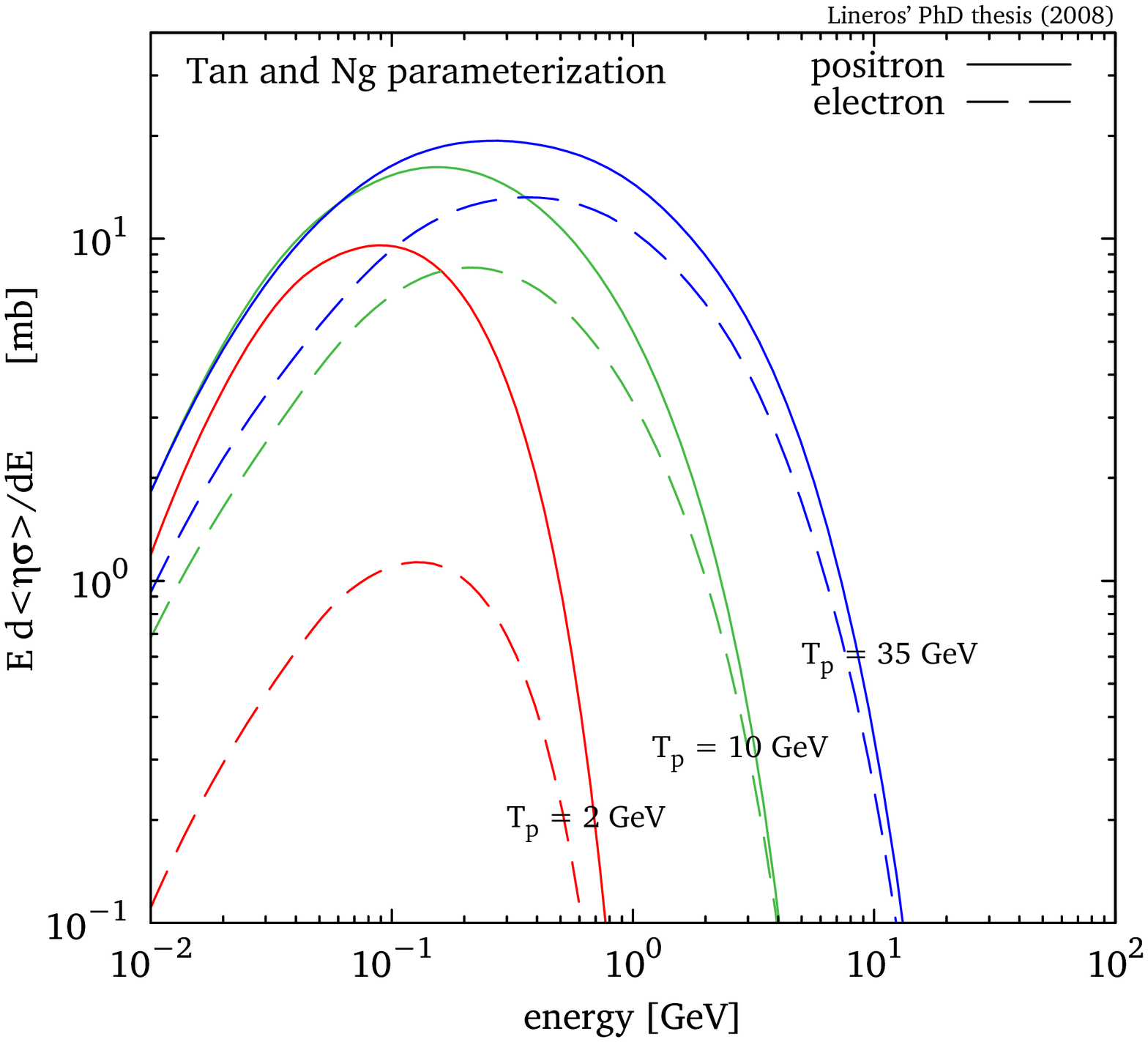}}\\
 \resizebox{\hsize}{!}{\includegraphics[angle=270, width=0.5\textwidth]{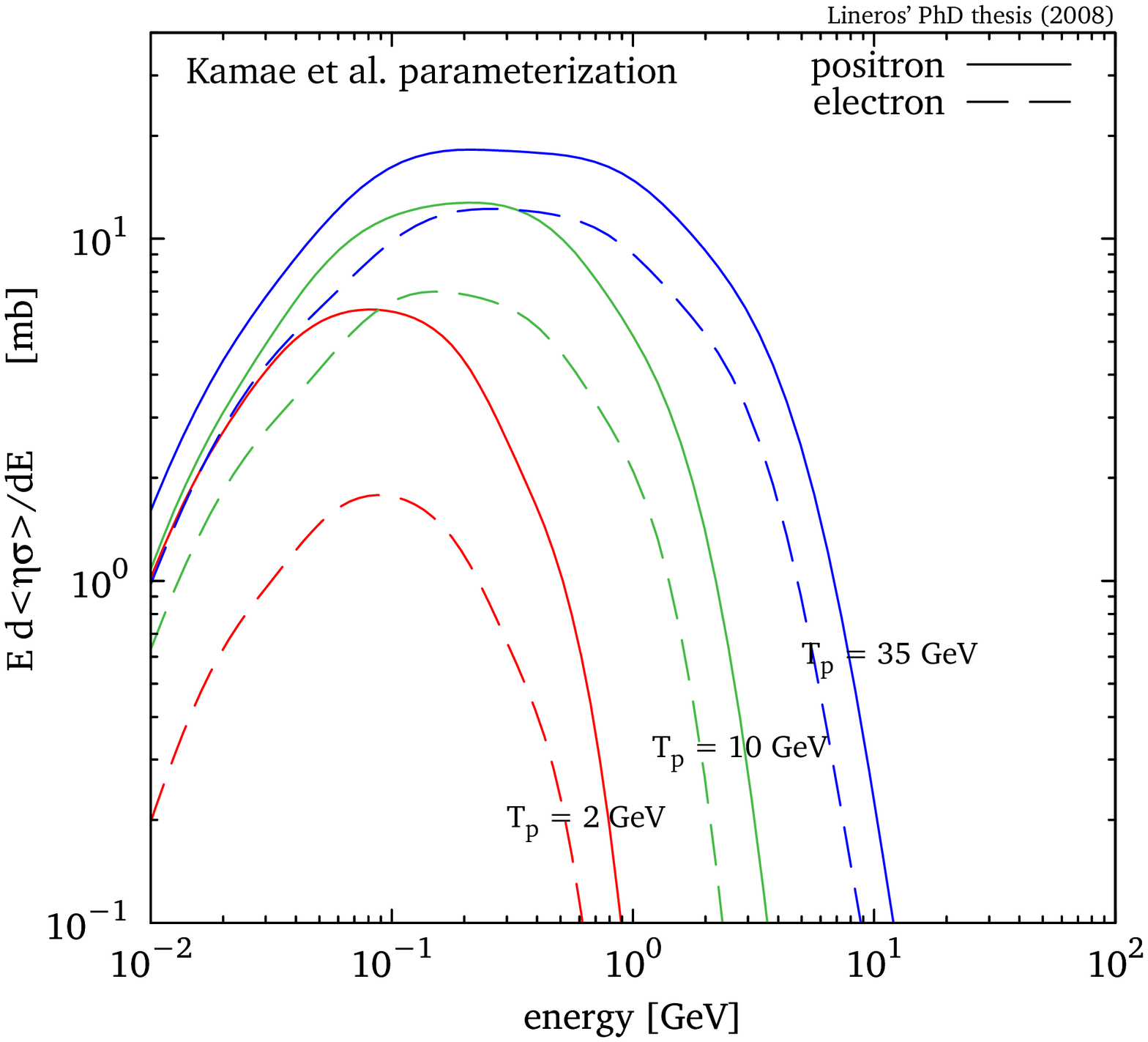} \includegraphics[angle=270, width=0.5\textwidth]{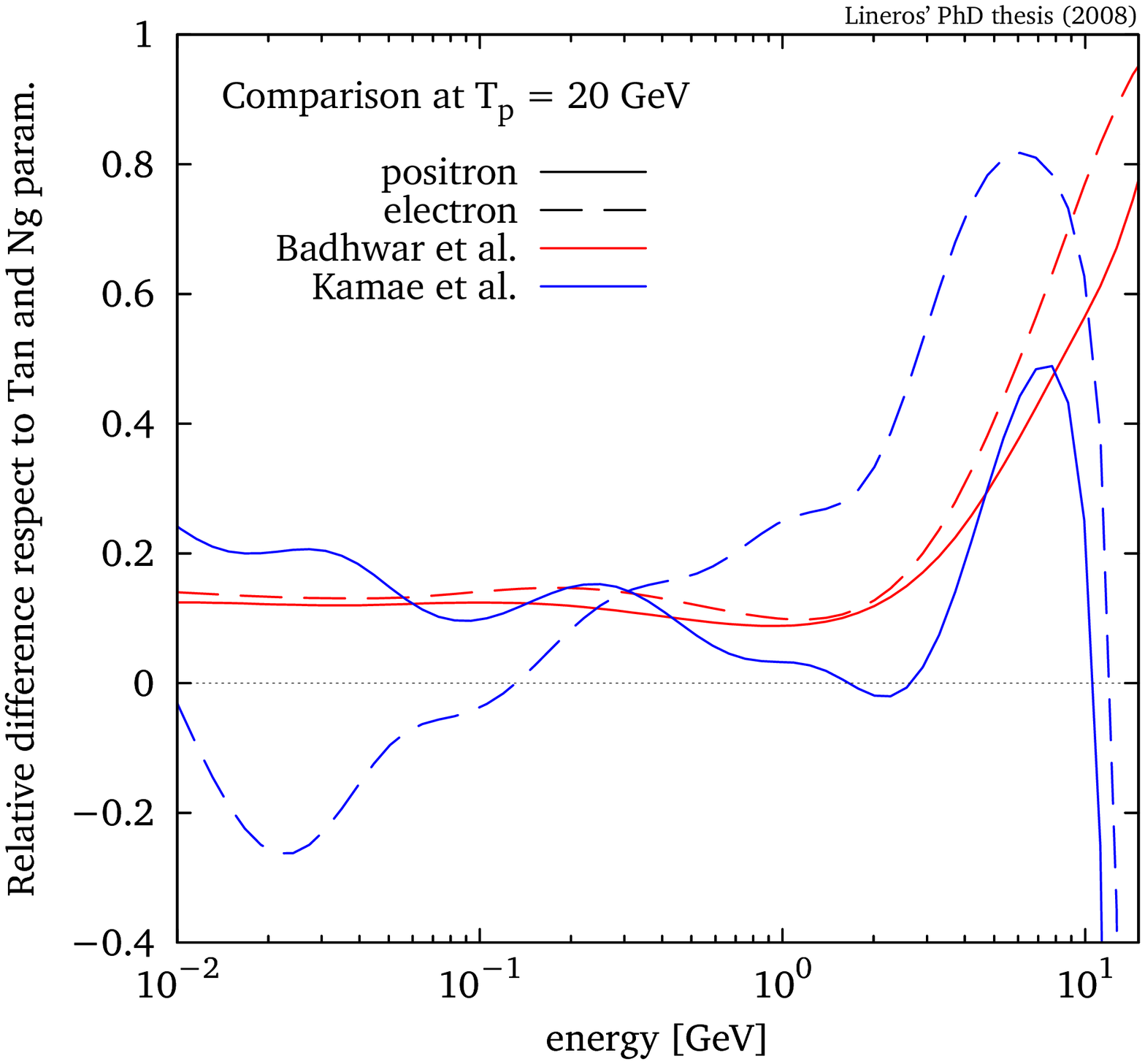}}
\caption{\label{f:crs-comp} Positron and electron inclusive cross section versus energy for proton energies of 2, 10 and 35 \tu{GeV}.
The different panels show the case of Badhwar et al.~\cite{Badhwar:1977zf}, Tan and Ng~\cite{Tan:1984ha}, and Kamae et al.~\cite{Kamae:2004xx} parametrizations. The last panel shows that the relative difference with respect to Tan and Ng solution is on average around 15\% and 25\% for Badhwar et al. and Kamae et al. cases respectively.}
\end{fig}
Previously we described different approaches to calculate the electron and positron spectra in proton--proton collisions.
Each parametrization is based on physical assumptions and encompasses available experimental data: 
therefore the small differences among them can be used to quantify the nuclear physics uncertainties.\\

In \citefig{f:crs-comp}, we show the comparisons among the three parametrizations for different proton kinetic energies.
We note that these parametrization are closely similar in behavior.
However, there are variations up to 80\% at proton energies of 20 \tu{GeV},  as in case of Kamae et al. versus Tan and Ng parametrization. Another feature is that the Kamae's parametrization estimates a smaller electron cross section with respect to the other two parametrizations.\\ 

Due to the low statistics at very low energy, Badhwar's and Tan's parametrization tend to produce non-physical distributions for proton kinetic energies below $6 \ \tu{GeV}$. Nevertheless, the total inclusive cross section -- the integrated version of those -- are still in agreement with the available experimental data. To fix this undesirables feature, both parametrization are patched by doing a smooth transition from 3 \tu{GeV} until 7 \tu{GeV} with the Stecker's model~\cite{Stecker:1967SA,Stecker:1970Ap}.
Let's clarify that Kamae's parametrization also includes that feature, but considering more resonances.\\
Moreover, for proton energies above 100~\tu{GeV}, Badhwar's parametrization becomes unstable specially the electron cross section case.\\

\subsection{\sc Production in annihilation processes}

Electron and positrons can also be the result of annihilations.
A first case to come in mind is the matter--antimatter annihilation like electron positron collisions at LEP.

The annihilation of dark matter also enters in this category, providing a new type of source of electrons and positrons that it would coexist with standard cosmic rays sources.\\

Independently of how these particles annihilate, electrons and positrons would come directly from the annihilation event (direct production case) or from the annihilation's sub-products, like decay of gauge or higgs bosons or hadronization/decay of quarks.\\

We work in a general approach, in which we generate electron and positron multiplicity distributions which are independent of dark matter physics and can be used for any purpose.\\

\subsubsection{\sc Multiplicity distribution}

In a generic annihilation case, the multiplicity distribution of electrons and positron is:
\begin{eq}
 \left(\frac{dn_e}{d\ener{}}\right)_{\chi \bar{\chi} \rightarrow e X} = \sum_{i} \tn{BR}\big(\chi \bar{\chi} \rightarrow i\big) \  \left(\frac{dn_e}{d\ener{}}\right)_{i \rightarrow e X} ,
\end{eq}

where we decompose the annihilation in intermediate states $i$, related directly to the annihilation mechanism via branching ratios, for instance, the dark matter annihilation case:
\begin{eq}
 \tn{BR}\big(\chi \bar{\chi} \rightarrow i\big) = \frac{\sigma\big(\chi \bar{\chi} \rightarrow i\big)}{\sigma_{\tn{total}}}.
\end{eq}
This presents a general decomposition based on bricks which are the multiplicity distribution for many states $i$.\\

\subsubsection{\sc Calculation of Multiplicity distributions}
\begin{tab}
	\begin{tabular}{|c|c|c|c|}
	\hline
	& \multicolumn{3}{c|}{Intermediate state} \\
	\hline \hline
	Charge & Leptons & Quarks & Gauge Bosons \\
	\hline
	+1 & $(\nue e^{+}) \ (\numu \mu^{+}) \ (\nutau \tau^{+})$ & $(u\bar{d})\ (c\bar{s})\ (t\bar{b})$ & $(\gamma W^{+})\ (Z W^{+})$ \\
	0 & $(e^{-} e^{+})\ (\mu^{-} \mu^{+})\ (\tau^{-} \tau^{+})$ & $(u\bar{u}) \ (d\bar{d}) \ (c\bar{c}) \ (s\bar{s}) \ (b\bar{b})\ (t\bar{t}) $ & $(gg) \ (ZZ) \ (W^{-} W^{+}) $\\ 
	-1 & $(e^{-} \nuebar) \ (\mu^{-} \numubar) \ (\tau^{-} \nutaubar )$ & $(d\bar{u})\ (s\bar{c})\ (b\bar{t})$ & $(W^{-} \gamma)\ (W^{-} Z)$ \\
	\hline
	\end{tabular}
	\caption{\label{t:md-pythia} Intermediate states for positron and electron multiplicity distributions.}
\end{tab}
We consider the most general set which is composed by pairs of Standard Model particles.
Each pair preserve quantum numbers of the annihilation, i.e. an electrically neutral and colorless final state.\\

Depending on the case, we calculate analytical distributions for simple cases (electrons, positrons, and muons),
and for complex ones, like the ones which involves hadronization processes, we used a modified version of PYTHIA~\cite{Sjostrand:2000wi} to generate the distribution, which includes the effect of polarization of muons.

The basic set is based on Standard Model particles~\citetab{t:md-pythia}.
Special cases like the Standard Model higgs and models like Two Higgs Doublet Models are easily composed using the basic set (details in~\cite{Lineros:2008hh}).\\

In~\citefig{f:md-pythia}, we present multiplicity distributions for cases inspired by annihilation of dark matter. 
We observe that the shape of each distribution depends directly on the intermediate state: states involving quarks produce a softer electron and positron spectra than leptonic cases.
In addition, the effect related to the polarization of muon in meson decays also has an impact, producing more energetic electrons than positrons.\\

\begin{fig}
	\resizebox{\hsize}{!}{\includegraphics[angle=270, width=0.5\textwidth]{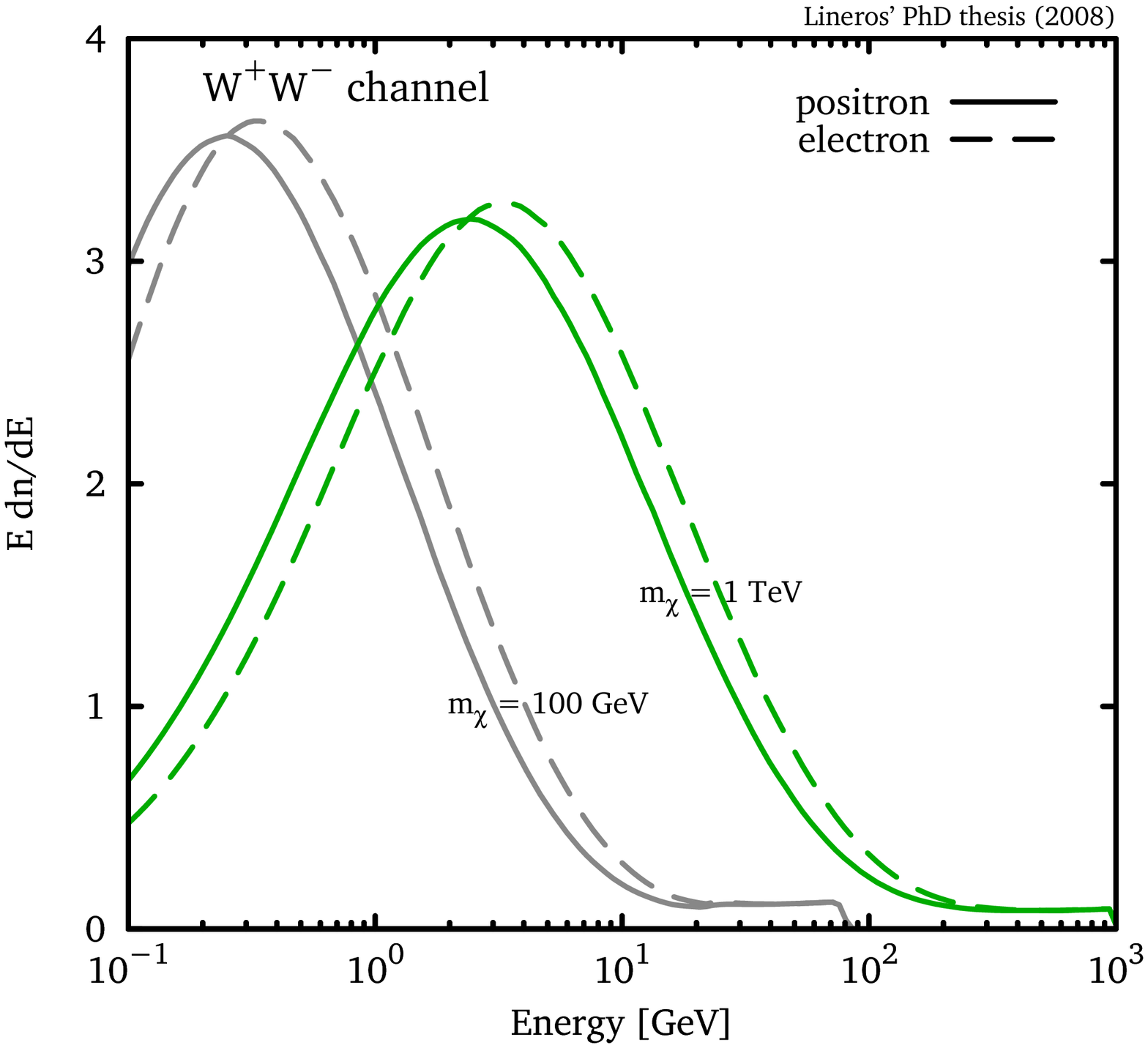}\includegraphics[angle=270, width=0.5\textwidth]{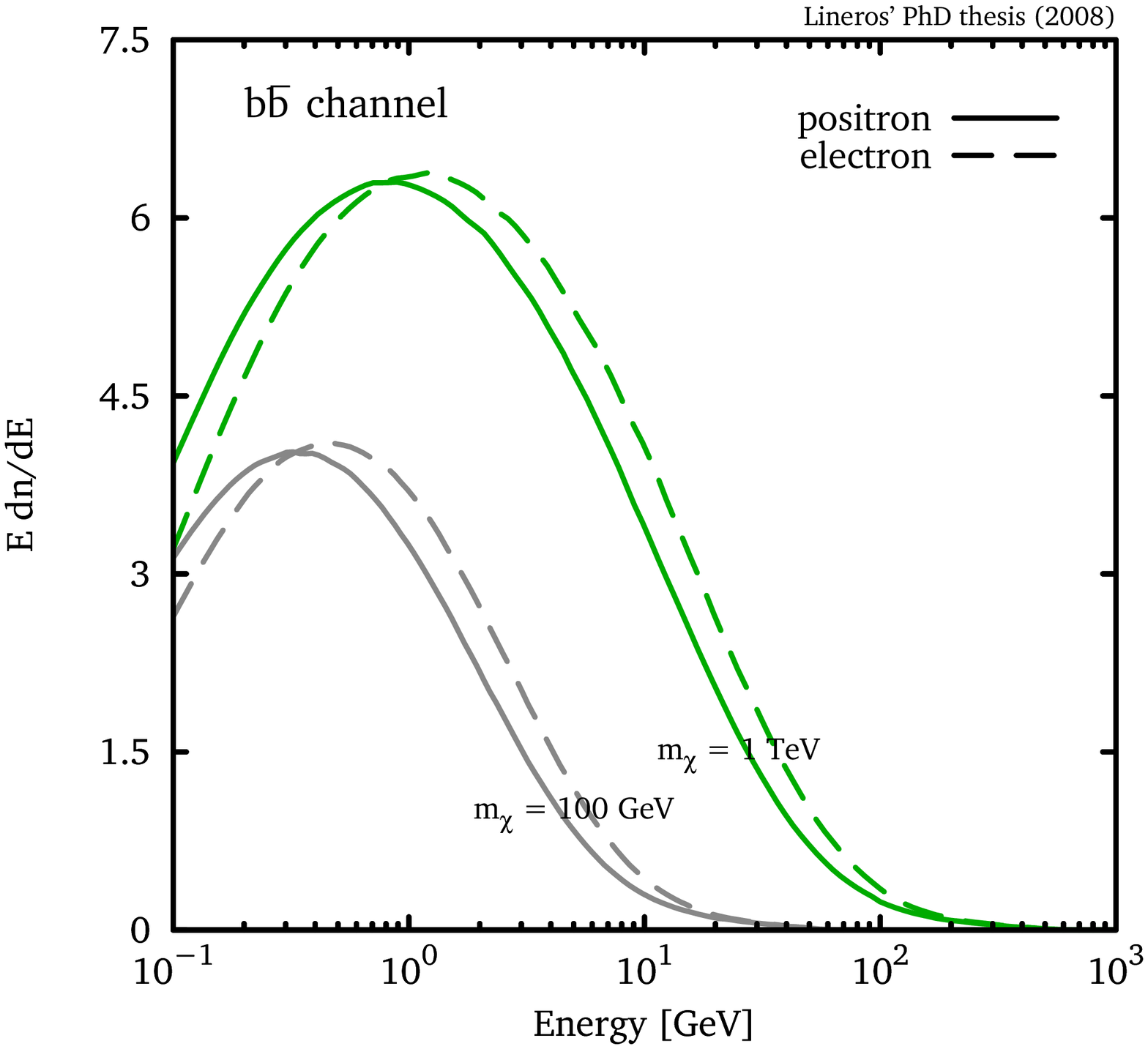}}
	\resizebox{\hsize}{!}{\includegraphics[angle=270, width=0.5\textwidth]{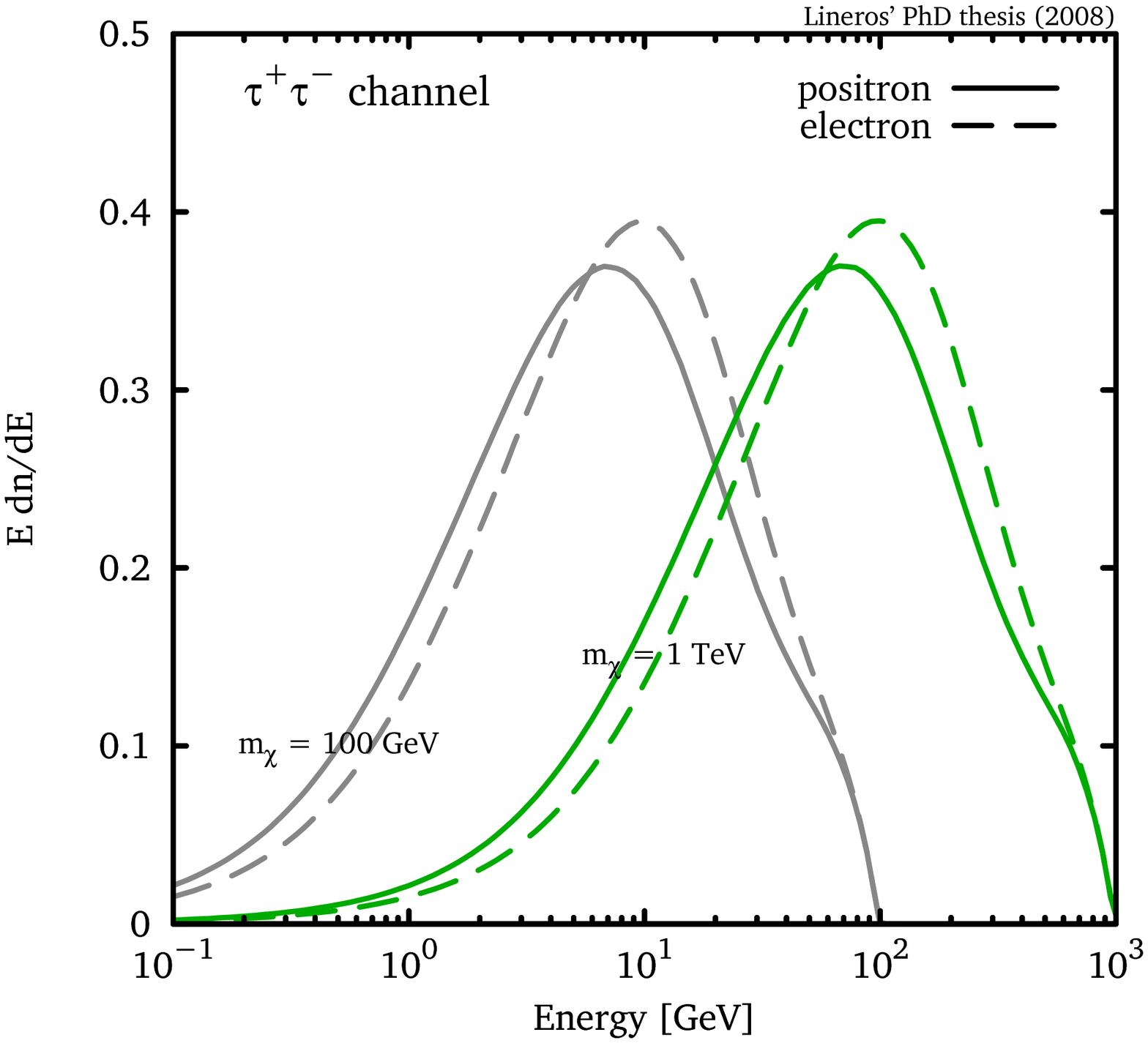}\includegraphics[angle=270, width=0.5\textwidth]{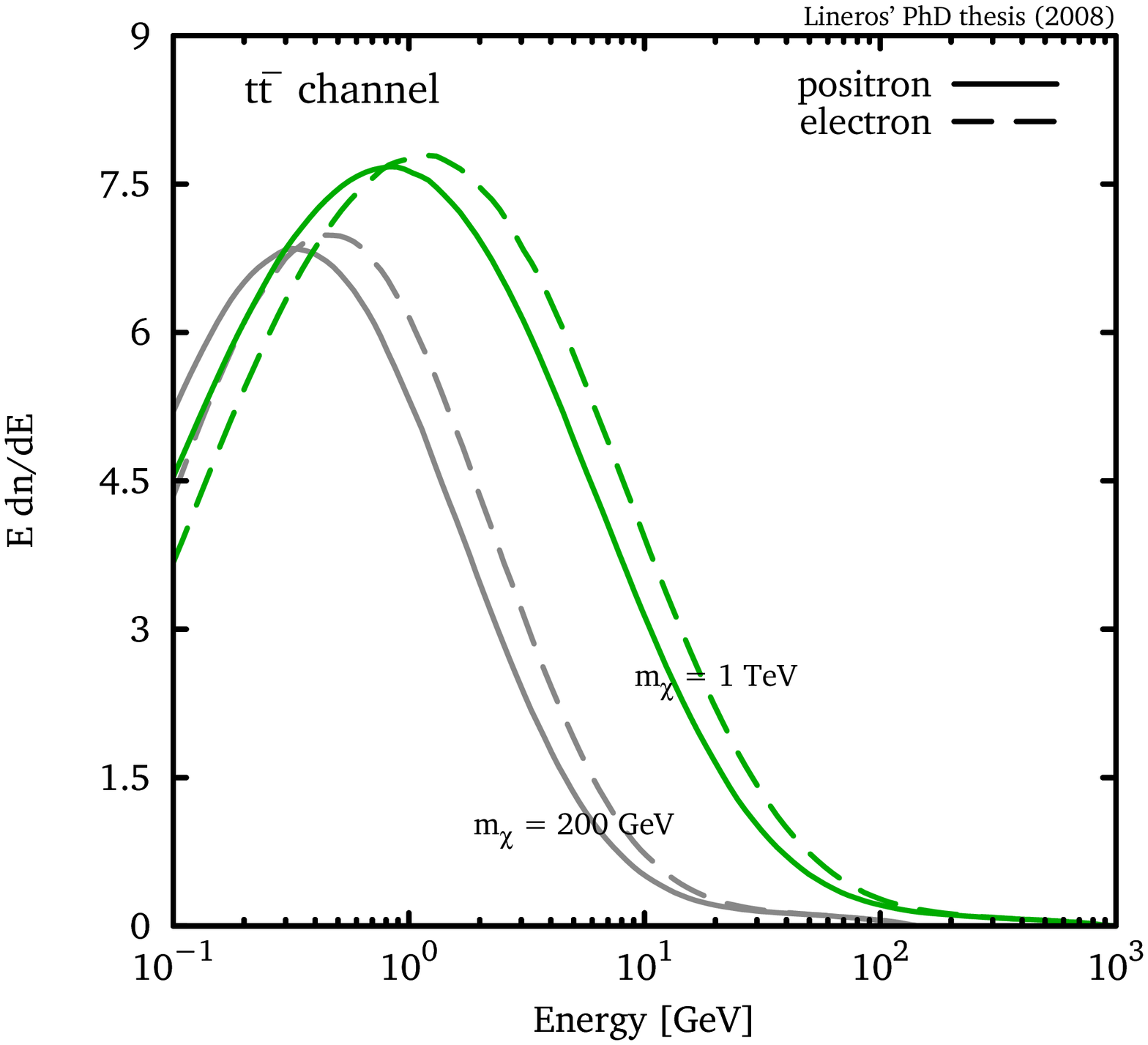}}
	\caption{\label{f:md-pythia} Multiplicity distribution for positrons and electrons versus energy. 
	Each panel shows a different intermediate state --  $W^{+}W^{-}$, $\tau^{+}\tau^{-}$, $b\bar{b}$, and $t\bar{t}$ --  with dark matter masses ($m_{\chi}$) of 100 \tu{GeV} (or 200 \tu{GeV}) and 1 \tu{TeV}.
	The different shapes on electron and positron distribution come from the effect of production of polarized muon.}

\end{fig}

%% file: tex/sect3.tex
\section{Propagation of positron and electron cosmic rays}
\label{sec:03}

The cosmic rays' journey from the source until the Earth is a complex problem.
Since cosmic rays start to travel, they are affected by many processes and their intensity depends on the cosmic ray energy scale. 
For instance, in the \tu{GeV}--range, their are strongly affected by turbulent magnetic fields that induce spatial diffusion, similar to the behavior of a single molecule in a gas.\\

For the GeV--scale cosmic rays, the Galactic medium has a very important role on the propagation.
Apart from the turbulent magnetic fields, there are also interactions with diffuse radiation fields (UV, IR, CMB). 
Their continuous interaction changes the cosmic rays energy, electrons and positrons are more affected than other species.\\
The combined effect of diffusion, energy losses, distance, among others, cause that the observed cosmic rays spectra is rather different from the original one. 

\subsection{\sc Two--Zone propagation model}
%
%
\begin{fig}
 \includegraphics[width=0.6\textwidth]{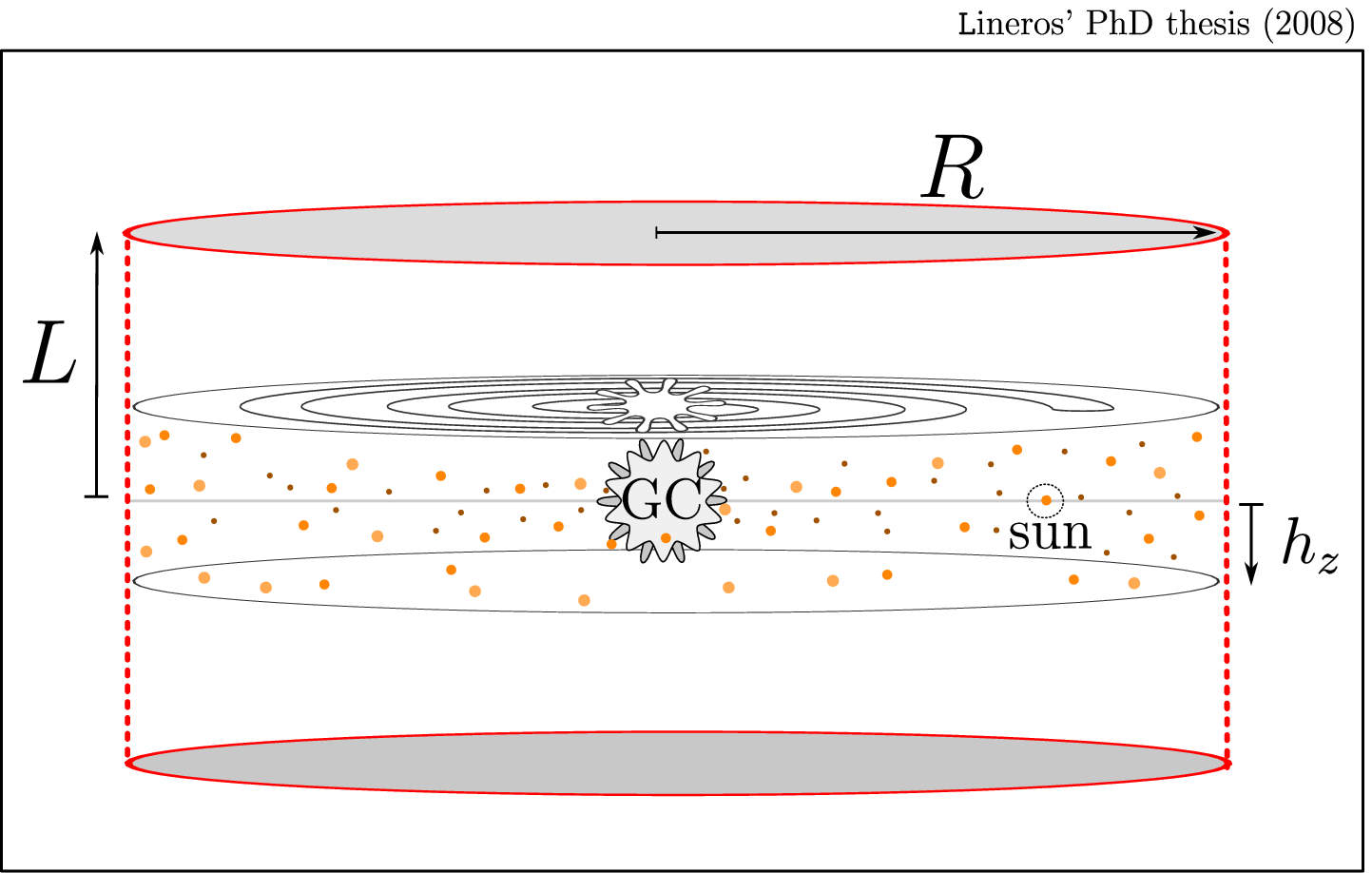}
 \caption{\label{f:pz-cyl} Propagation Zone geometry for the Milky Way. A cylinder of radius R ($= 20\;\tu{kpc}$) with thickness of $2\;L$ delimits the region where cosmic rays propagate. A small cylinder with same radius but with thickness $2\;h_z$ ($= 200\;\tu{pc}$) models the Galactic plane. The Solar system is placed at the Galactic plane with distance $r_{\astrosun} = 8.5 \; \tu{kpc}$ from the Galactic center.}
\end{fig}
The modelization of cosmic rays propagation is a hard task due to the propagation complexity.
We use a successfully tested model for propagation of cosmic rays~\cite{Ginzburg:1980, Maurin:2001sj} which is tuned properly to explain the current observations on abundances of nuclear cosmic rays.
In general terms, this model is divided in two parts:
\begin{itemize}
\item The propagation zone: 
the region related to the extension of turbulent magnetic fields, where cosmic rays propagate in a diffusive regime.
Outside this region, we suppose cosmic rays propagate in straight lines: this induces cosmic rays leaking from the diffusive zone.\\
The propagation zone is composed by two cylinders centered at the Galactic center (\citefig{f:pz-cyl}). 
Both cylinders have a common radius equal to the galactic one ($R = 20\;\tu{kpc}$).
The \emph{thick} cylinder has a height of $2L$ and fills the whole propagation zone. 
Its height is constrained by measurements of many species of cosmic rays~\cite{Maurin:2001sj}.
The second cylinder is a \emph{thin disk} with height $2 h_z$, where $h_z = 100\;\tu{pc}$.
It also contains the interstellar medium, cosmic rays sources and interactions among them.\\

\item The transport equation: 
it corresponds to a continuity equation for the number density of cosmic rays per unit of energy.
It contains all physical processes that are related to cosmic rays physics, like energy losses, diffusion, re-acceleration and cosmic rays sources.\\
\end{itemize}

\subsection{\sc Transport Equation for electrons and positrons}
In this case, the transport equation is slightly different from the one for nuclei cosmic rays because the dominant energy loss process is the interaction with radiation fields (via inverse Compton scattering) which is much efficient in these particles than in nuclei cosmic ray.\\
Then the transport equation for number density of electron and positrons ($\psi$) is:
\begin{eq}
	\label{eq:std_te}
	\frac{\partial \psi}{\partial t}  - \nabla \big( D(\ener{}) \nabla \psi \big) - \frac{\partial}{\partial \ener{}} \big(b(\ener{}) \psi \big) = q \, ,
\end{eq}
where $D(\ener{})$ is the diffusion coefficient, $b(\ener{})$ the energy loss term, and $q$ the source term.\\
The diffusion term is considered homogeneous in space with a energy dependence:
\begin{eq}
	D(\ener{}) = K_0 \, \left(\frac{\ener{}}{\ener{0}}\right)^{\delta} \, ,
\end{eq}
where $K_0$ and $\delta$ are phenomenological parameters inspired by models of the interstellar medium based on magneto hydrodynamics. 
$\ener{0}$ is a normalization energy scale, here fixed at the value of 1~\tu{GeV}.\\

Energy losses are related to interactions of electron and positrons with the interstellar radiation fields, the cosmic microwave background, and the galactic magnetic field.
Commonly, this term is:
\begin{eq}
	\label{e:eloss-std}
	b(\ener{}) = \frac{\ener{0}}{\tau_E} \, \left(\frac{\ener{}}{\ener{0}}\right)^2 \, ,
\end{eq}
where $\tau_E( = 10^{16} \, \tu{s})$ corresponds to an effective energy loss time calculated via the inverse Compton scattering with the radiation fields in the Thomson regime.
However, this term is no longer valid for energies larger than  $\mass{e}^2/\langle\ener{\gamma}\rangle$, where $\langle\ener{\gamma}\rangle$ is the radiation field energy,  and it has to be corrected considering the Klein-Nishina cross section~\cite{klein:1929,1970RvMP...42..237B,1971PhRvD...3.2308B}.\\

%
%
\begin{fig}
 \includegraphics[width=0.6\textwidth]{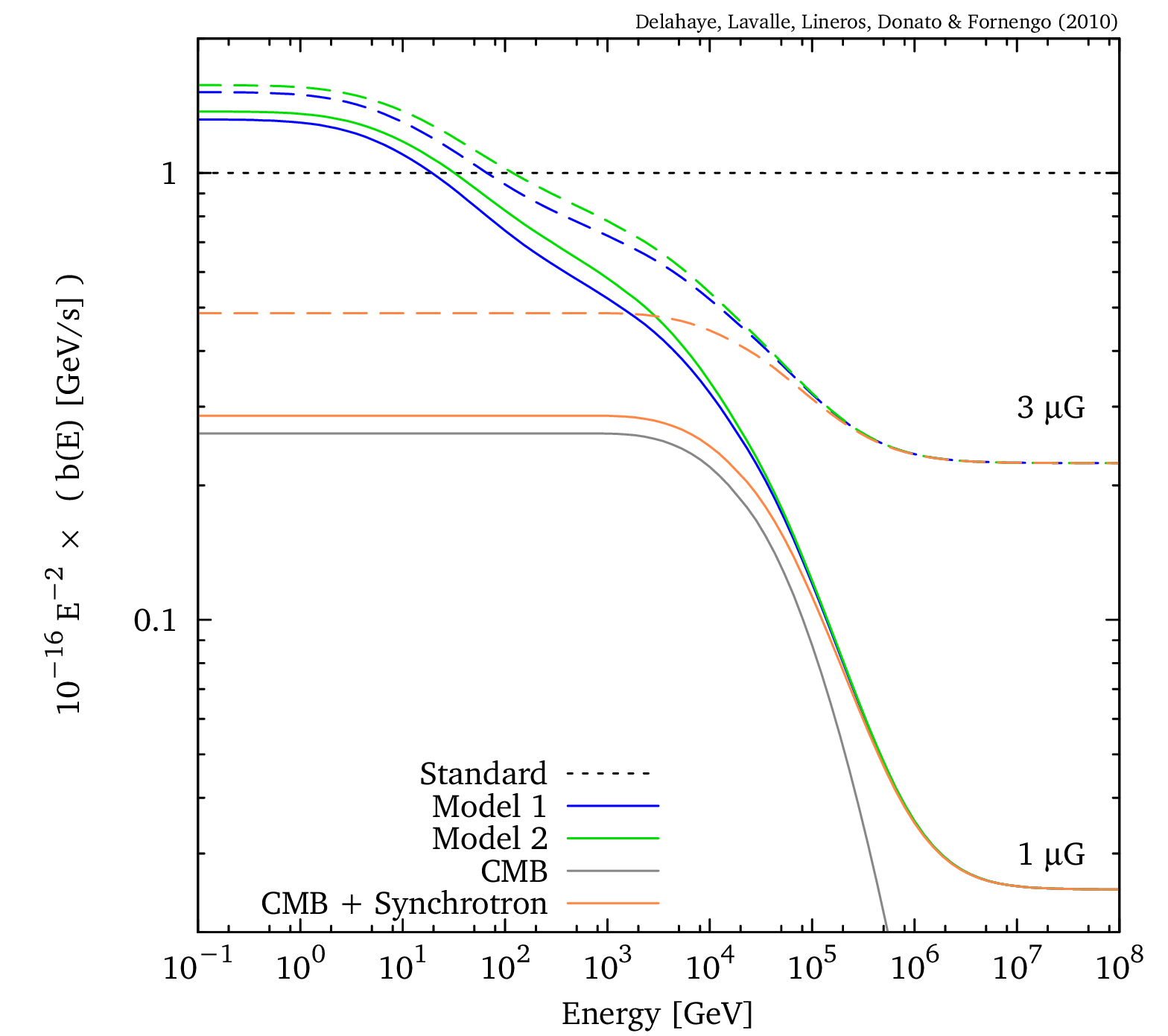}
 \caption{\label{f:eloss-mol} Energy loss term $10^{16} \ener{}^{-2} b(\ener{})$ versus electron (positron) energy. The energy loss for different radiation fields models and the standard assumption are shown}
\end{fig}
In~\citefig{f:eloss-mol}, we present the energy loss term for the different components of the radiation field and those are compared to the standard assumption of the Thomson regime~(\citeeq{e:eloss-std}). The figure is from our recent work~\cite{Delahaye:2010ji} that follows the line of the Author's thesis.\\

\subsection{\sc Transport equation solution}
We solve the differential eaution by means of Green functions which correspond to the solution for a point--like source:
\begin{eq}
	q(t,\vec{x},\ener{}) = \delta(t-t_s) \delta(\ener{} - \ener{s}) \delta^3(\vec{x} - \vec{x}_s) \, ,
\end{eq}
where $t_s$, $\ener{s}$, and $\vec{x}_s$ correspond to the injection time, energy, and position.\\

The green function is easily obtained through the Fourier transformation method~\cite{Baltz:1998xv,Lineros:2008hh} and it has an analytical form~\cite{Delahaye:2010ji}:
\begin{eq}
	G(t,t_0,\vec{x},\vec{x}_s,\ener{}, \ener{s})  &=& \delta(t-t_s-\tau_c) \, \frac{\widetilde{G}(\vec{x},\vec{x}_s,\lambda_d)}{b(\ener{})} \\ &=& \delta(t-t_s-\tau_c) \, G_{\tn{ss}}(\vec{x},\vec{x}_s,\ener{}, \ener{s})\, , \nonumber
\end{eq}
where $\widetilde{G}$ is the \emph{tilded green function}: 
\begin{eq}
	\label{eq:free-space-g}
	\widetilde{G}(\vec{x},\vec{x}_0,\lambda_d) = \frac{1}{\pi^{3/2} \lambda_d^3} \exp\left(-\frac{(\vec{x}-\vec{x}_0)^2}{\lambda_d^2}\right) \, ,
\end{eq}
and it is proportional to the Green function $(G_{\tn{ss}})$ in the time independent case (steady-state regime).\\

The terms $\tau_c$ and $\lambda_d$ are rich in physical meaning and come naturally from the solution.
$\tau_c$ is the \emph{cooling time} which corresponds to the time lapses for an electron or positron to reduce its energy from $\ener{s}$ to $\ener{}$:
\begin{eq}
	\tau_c(\ener{},\ener{s}) = \int_{\ener{}}^{\ener{s}} d\epsilon \; 1/b(\epsilon) \, .
\end{eq}
In the same way, $\lambda_d$ is the \emph{diffusion length} which gives a scale for the propagation and is defined by:
\begin{eq}
	\lambda^2_d(\ener{},\ener{s}) = 4 \int_{\ener{}}^{\ener{s}} d\epsilon \; D(\epsilon)/b(\epsilon) \, .
\end{eq}
The former two quantities become fundamental to describe the propagation. Also those can be calculated for any type of diffusion coefficient and energy loss term allowing us to explore new forms, like: deviations from the standard diffusion coefficient and extra energy loss processes.\\


%
With the Green function, we are able to find the solution for the transport equation for any generic source $Q$.
The solution is found by convoluting the source with the green function in the following way:
\begin{eq}
	\psi(t,\vec{x},\ener{}) = \int_{-\infty}^{t} dt_s  \int_{\ener{}}^{\infty} d\ener{s} \int d^3 x_s \; G(t,t_s,\vec{x},\vec{x}_s,\ener{}, \ener{s}) \times Q(t_s, \vec{x}_s, \ener{s}) \, ,
\end{eq}
where the integration limits are set to consider only the physical contribution from the source $Q$.\\

As we previously said, the model considers that cosmic rays may escape from the propagation zone when they reach its boundaries.
This means:
\begin{eq}
 z=\pm L \vee \sqrt{x^2 + y^2} = R \ \Longrightarrow \ \psi(t,\vec{x},\ener{}) = 0 \, ,
\end{eq}
these conditions just affect the tilded Green function (\citeeq{eq:free-space-g}).\\

There are two standard approaches to impose the boundary conditions:
\begin{itemize}
	\item Eigenfunction expansion: 
The transport equation is solved using a complete set of Helmholtz equation eigenfunctions~($\chi_g$) that naturally respect the boundary conditions.
The green function is analytical in some cases and correspond to the sum on eigenvalues~($g$):
\begin{eq}
	\widetilde{G}_{\tn{bc}}(\vec{x},\vec{x}_s,\lambda_d) = \sum_g \chi_g^{\dagger}(\vec{x}_s) \chi_g(\vec{x}) \exp\left( - \frac{1}{4} g^2 \lambda_d^2\right)  \, .
\end{eq}
We use the Fourier--Bessel expansion, where harmonics functions are used for vertical coordinate and first kind Bessel functions for the radial one.
The method is very well behaved and fast to calculate when $\lambda_d$ is rather big~\cite{Delahaye:2007fr, Delahaye:2008, Lineros:2008hh}.\\

	\item Method of Images:
Also known as Method of Inversion, it consists on adding counter terms or mirror sources that compensate the effect of the original source, preserving the boundary condition.\\

For the vertical coordinate, we include an infinite number of mirror sources transforming the green function into a sum:
\begin{eq}
	\widetilde{G}_{\tn{vertical bc}}(\vec{x},\vec{x}_s,\lambda_d) = \sum_{n=-\infty}^{\infty} (-1)^n \; \widetilde{G}(\vec{x},\vec{x}_{s,n},\lambda_d) \,,
\end{eq}
where $\vec{x}_{s,n} = (x_s\; , y_s\; , 2Ln + (-1)^n z_s)$ and $\widetilde{G}$ is the free space tilded green function~(\citeeq{eq:free-space-g}).
This form is very efficient for small values of $\lambda_d$ and complements the expansion in harmonic functions.\\ 

The Green function that respects radial boundary conditions is obtained by add just one extra counter term:
\begin{eq}
	\widetilde{G}_{\tn{radial bc}}(\vec{x},\vec{x}_s,\lambda_d) = \widetilde{G}(\vec{x},\vec{x}_{s},\lambda_d) - \widetilde{G}(\vec{\sigma},\vec{\sigma}_{s},\lambda_d) \,
\end{eq}
where $\vec{\sigma} = (\beta x, \beta y, z)$,  $\displaystyle \vec{\sigma}_s = (\frac{x_s}{\beta}, \frac{y_s}{\beta}, z_s)$ and $\displaystyle \beta^2 = \frac{x_s^2 + y_s^2}{R}$.\\
We remark that this Green function works for all values of $\lambda_d$ and does not need any kind of special symmetry like the Bessel expansion. 
\end{itemize}

\subsection{\sc Space of Parameters of the model}

The two--zone propagation model is also used for studying nuclei cosmic rays. 
Previous studies on the ratio Boron and Carbon (B/C) constraint the space of parameters to a volume which is fully compatible with current observations in many other species.
We use the allowed space of parameters to study positron and electron cosmic rays consistently and systematically with respect to cosmic rays propagation (details in \cite{Maurin:2001sj,Maurin:2002uc}).\\

%% file: tex/sect4.tex
\section{Positron cosmic rays from dark matter annihilation}
\label{sec:04}

Secondary positrons and electrons are produced in the Milky Way from the interaction of nuclei cosmic rays on the interstellar gas~\cite{Moskalenko:1997gh} and are an important tool for the comprehension of cosmic-ray propagation~\cite{Maurin:2001sj}.
Data on the cosmic positron flux (often reported in terms of the positron fraction) have been collected by several experiments \cite{Barwick:1997ig,Ahlen:1994ct,Alcaraz:2000PhLB,Aguilar:2007,Boezio:2000,Grimani:2002yz}. \\

We point out the HEAT balloon experiment~\cite{Barwick:1997ig} that has mildly indicated a possible positron excess at energies larger than 10~\tu{GeV} with respect to the current -- at that moment -- calculations for the secondary component~\cite{Moskalenko:1997gh}.
In October 2008, the latest results of PAMELA experiment~\cite{Boezio:2004jx} have confirmed and extended this feature~\cite{Adriani:2008zr}. \\

Different astrophysical contributions to the positron fraction in the 10~\tu{GeV} region have been explored \cite{Barwick:1997ig}, but only accurate and energy extended data could confirm the presence of a bump in the positron fraction and its physical interpretation. 
Alternatively, it has been conjectured that the possible excess of positrons found in the HEAT data could be due to annihilation of dark matter in the galactic halo~\cite{Baltz:1998xv,Hooper:2004bq}.
Although, this interpretation is limited by uncertainties in the halo structure and in the cosmic rays propagation modeling.\\

This section is based on our work~\cite{Delahaye:2007fr}. 
We study the propagation of the positrons related to dark matter annihilations in connection with the study of the uncertainties due to propagation models compatible with B/C measurements~\cite{Maurin:2001sj}.\\

\subsection{\sc dark matter annihilation like source of positrons}

%
%
\begin{tab}
\begin{tabular}{|l|cccc|}
\hline
Halo model & $\alpha$ & $\beta$ & $\gamma$ & $r_s$ [\tu{kpc}] \\
\hline 
Cored isothermal~\cite{Bahcall:1980fb}
&  2  &  2  &  0  &  5  \\
Navarro, Frenk \& White~\cite{Navarro:1996gj}
&        1        &        3        &        1        &        20       \\
Moore~\cite{Diemand:2004wh}
&        1.5      &        3        &        1.3      &        30       \\
\hline
\end{tabular}
\caption{\label{tab:indices}Dark matter distribution profiles in the Milky Way.}
\end{tab}
According to the various supersymmetric theories, the annihilation of a dark matter particles points to the direct production of an electron-positron pair or to the production of many species subsequently decaying into photons, neutrinos, hadrons and positrons. \\
In our study, we consider four possible annihilation channels which would appear in any model of dark matter.
The first corresponds to direct production of a $e^+ e^-$ pair and it is better motivated in theories with universal extra-dimension~\cite{Cheng:2002iz,Servant:2002aq,Appelquist:2000nn}.
We alternatively consider annihilations into $W^{+} W^{-}$, $\tau^+ \tau^-$ and $b\bar{b}$ pairs.\\

For any annihilation channel, the source term is written as:
\begin{eq}\label{source}
q_{\tn{dm}}\left(\vec{x},\ener{e}\right) = \eta \; \langle\sigma v\rangle \; \left( {\displaystyle \frac{\rho(\vec{x})}{\mass{\chi}}} \right)^{2} \; \dnde{e} \; ,
\end{eq}
where the $\eta$ is a quantum coefficient which depends on whether the particle is or not its own antiparticle.
$\langle\sigma v\rangle$ corresponds to the thermally averaged annihilation cross section, its value depends on the specific supersymmetric model and it is also constrained by cosmology~\cite{Arina:2007tm,Arina:2007}. 
We have actually taken here a benchmark value of $2.1 \times 10^{-26}$~\tu{cm}$^{3}$~\tu{sec}$^{-1}$ which leads to a relic abundance of $\Omega_{\chi} h^{2} \sim 0.14$ (in agreement with the WMAP observations~\cite{Spergel:2006hy,Hinshaw:2008kr}).
The dark matter mass value ($m_{\chi}$) is still unknown but for the case of neutralinos, theoretical arguments as well as the LEP and WMAP results constrain its mass to range from a few \tu{GeV}~\cite{Bottino:2002ry,Bottino:2003iu,Belanger:2002nr,Hooper:2002nq} up to a few \tu{TeV}. 
Keeping in mind the positron excess, we explore the cases with a dark matter mass of 100~\tu{GeV} and 500~\tu{GeV}.\\

%
$dn/d\ener{e}$ represents the multiplicity distribution of electrons (positrons) per single annihilation (details in~\cite{Lineros:2008hh}).

%
The astronomical ingredient on the source term~(\citeeq{source}) is the dark matter distribution $\rho(\mb{x})$ inside the Milky Way halo. 
We have considered the generic profile:
\begin{eq} 
\label{eq:profile} 
\rho(r) = \rho_{\astrosun} \; \left( {\displaystyle \frac{r_{\astrosun}}{r}} \right)^{\gamma} \; \left( {\displaystyle \frac{1 \, + \, (r_{\astrosun}/r_{s})^{\alpha}} {1 \, + \, (r/r_{s})^{\alpha}}} \right)^{(\beta - \gamma) / \alpha} \;\; , 
\end{eq} 
where $r_{\astrosun} = 8.5$~\tu{kpc} is the galactocentric distance of the Solar System. 
Note that $r$ denotes the radius in spherical coordinates.
The local dark matter density has been set equal to $\rho_{\astrosun} = 0.3$~\tu{GeV}~\tu{cm}$^{-3}$~\cite{Berezinsky:1992}.
We discussed three profiles: an isothermal cored distribution~\cite{Bahcall:1980fb} for which $r_{s}$ is the radius of the central core, the Navarro, Frenk and White profile~\cite{Navarro:1996gj} (hereafter NFW) and Moore's model~\cite{Diemand:2004wh}. 
%
%
However, some studies point out to that the central cusp may be less favorable in spiral galaxies and proposed a universal dark matter distribution~\cite{Gentile:2004tb, Salucci:2007tm}.
The NFW and Moore profiles have been numerically established thanks to N-body simulations.
In the case of the Moore profile, the index
$\gamma$ lies between 1 and 1.5 and we have chosen a value of 1.3~(\citetab{tab:indices}) which is more representative.\\

The possible presence of dark matter substructures inside these smooth distributions enhances the annihilation signal by the so--called boost factor, although the boost factor value is still open to debate~\cite{Baltz:2001ir,Kane:2001fz,Jeannerot:1999yn}.
%
It has recently been shown that the boost factor due to substructures in the dark matter halo depends on the positron energy and on the statistical properties of the dark matter distribution~\cite{Lavalle:2006vb}.
In addition, it has been pointed out that its numerical values is quite modest~\cite{Lavalle:1900wn}, being of the order of 10--30.\\

\subsection{\sc the positron flux and fraction}

%
\begin{fig}
\includegraphics[angle=270, width=0.8\textwidth]{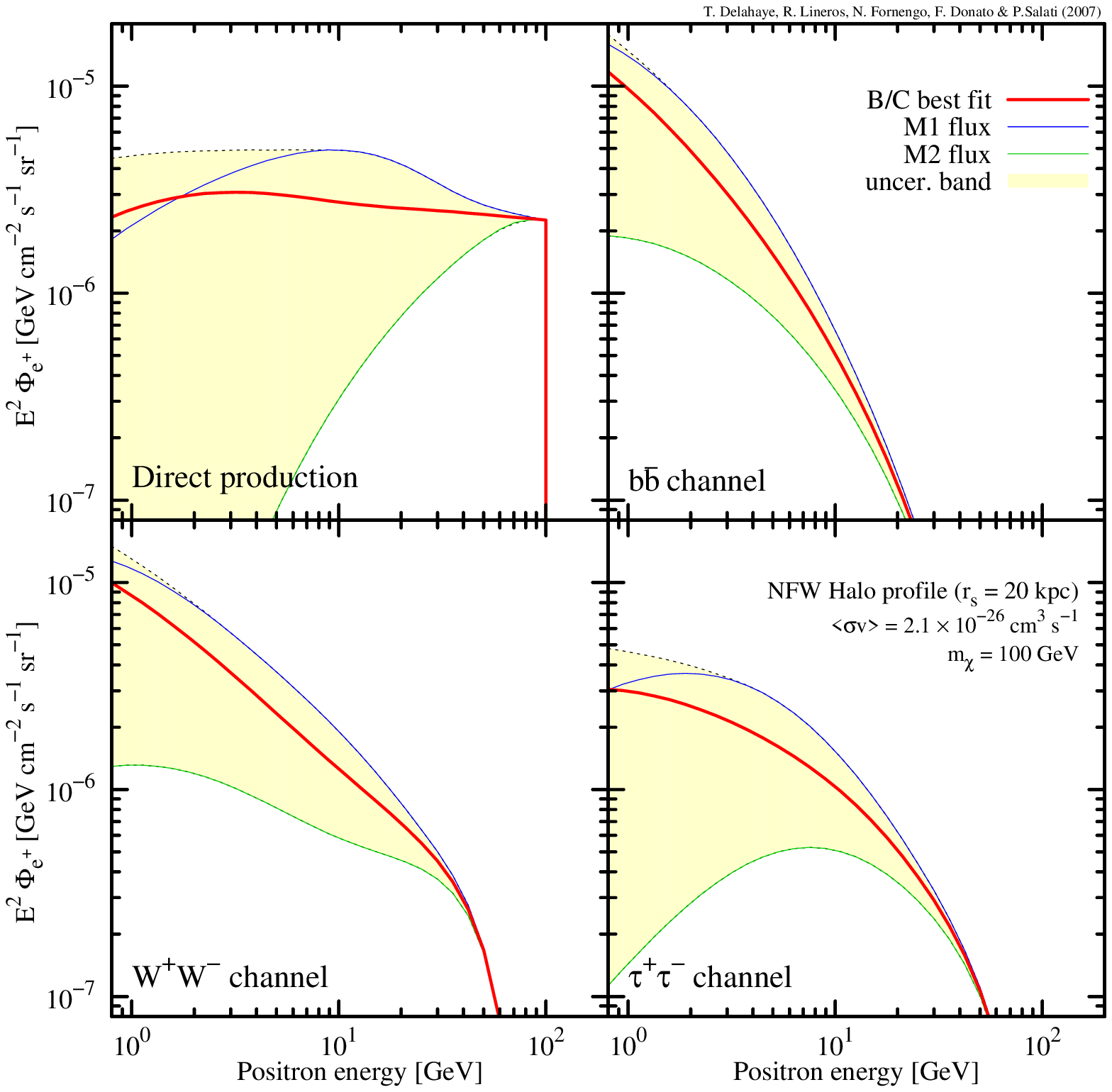}
\caption{ \label{fig:f1-100gev} Positron flux $E^2 \Phi_{e^+}$ versus the positron energy $\ener{}$, for a dark matter particle with a mass of 100~\tu{GeV} and for a NFW profile~(\citetab{tab:indices}).
The four panels refer to different annihilation final states: direct $e^+ e^-$ production (top left), $b\bar{b}$ (top right), $W^+ W^-$ (bottom left) and $\tau^+ \tau^-$ (bottom right).
In each panel, the thick solid [red] curve refers to the best--fit choice (MED) of the astrophysical parameters. The upper [blue] and lower [green] thin solid lines correspond respectively to the astrophysical configurations which provide here the maximal (M1) and minimal (M2) flux. The numerical values of these configuration are defined in~\cite{Delahaye:2007fr}.
The colored [yellow] area features the total uncertainty band arising from positron propagation.}
\end{fig}
%

Using the source term~(\citeeq{source}), we calculate the propagated positrons flux for many cases (\citefig{fig:f1-100gev}):
\begin{itemize}
\item Annihilation channels: The nature of dark matter fixes somehow the annihilation channel.
We observe how channels -- like the direct production one -- produce harder spectra than channels involving quarks  because positron are produced with energy equal to dark matter particle mass.
This is directly related to the multiplicity distributions.
Positrons in quark--antiquark channels come from hadronization processes and produce softer spectra with behavior similar to a power law: $\ener{e}^{-3 \pm 0.8}$.
On the other hand, channels involving muons or gauge boson $W^{\pm}$ produce harder spectra since positrons are produced at earlier stages, in decay chains of the original particles, taking a big fraction of the available energy.\\
\item Propagation uncertainties: 
Different propagation scenarios have different impact on positron generated by dark matter annihilation.
In \citefig{fig:f1-100gev}, we present the uncertainty band associated to the B/C analysis. 
We notice that the size of the band depends on the annihilation channel.
The direct production case presents the largest uncertainty band at low energies because the low energy flux is the result of far-away propagated positrons and not from the very local ones.
This also explains why at energies closer to 100~\tu{GeV}, the uncertainty band is smaller converges to an unique curve.
Other channels are less affected because very local produced positrons are not affected by the propagation parameters.
\end{itemize}

Let us stress that the positron flux obtained from annihilation of dark matter is not the only one.
We need to consider other astrophysical components. 
Due to the nature of astrophysical processes, positrons are dominated by a secondary component, i.e.  those are created from the interaction of nuclei cosmic rays with the interstellar gas.\\

To study the behavior of the positron signal, we include the secondary positron component and the electron flux from parametrized fluxes~\cite{Baltz:1998xv}.\\
In \citefig{fig:f3-heat-pf-100gev} we present the effect of the annihilation channel and propagation uncertainties on the positron fraction.
We observe how channels with harder spectra are more suitable to explain the positron excess. 
The $b\bar{b}$ case is less favorable because most of the positrons are at low energy, making impossible to reproduce the observations.\\
The uncertainties of propagation are sizable with respect to the signal from dark matter, however the impact is not enough to destroy some features arising from the annihilation channels.\\
%

%
%

%

%
%
\begin{fig}
\includegraphics[angle=270, width=0.8\textwidth]{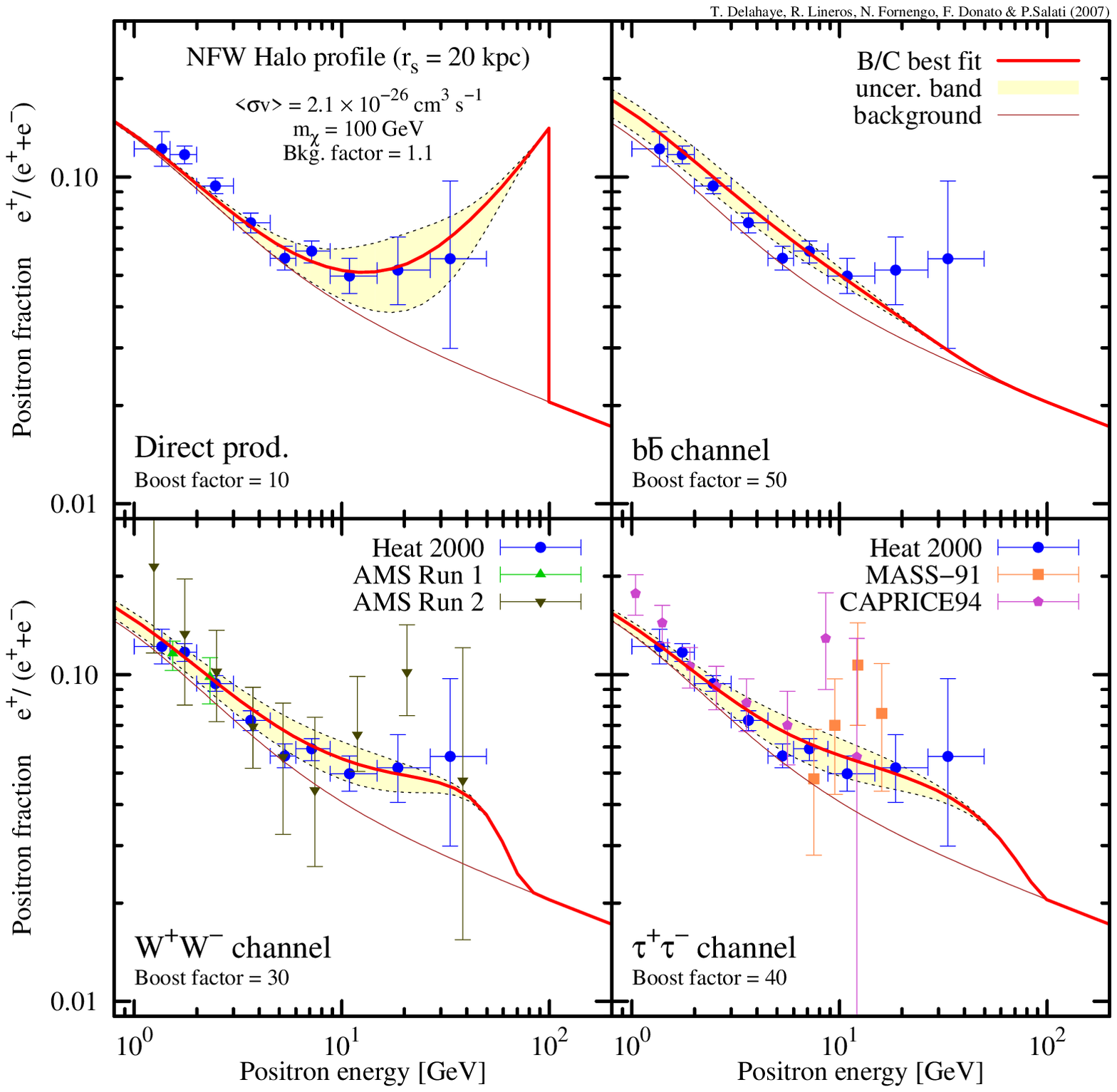}
\caption{\label{fig:f3-heat-pf-100gev} Positron fraction $e^+/(e^- + e^+)$ versus the positron detection energy $\ener{}$. Notations are as in \citefig{fig:f1-100gev}. In each panel, the thin [brown] solid line stands for the background \cite{Baltz:1998xv, Moskalenko:1997gh} whereas the thick solid [red] curve refers to the total positron flux where the signal is calculated with the best--fit choice (MED) of the astrophysical parameters.
Experimental data from HEAT~\cite{Barwick:1997ig}, AMS~\cite{Alcaraz:2000PhLB,Aguilar:2007}, CAPRICE~\cite{Boezio:2000} and MASS~\cite{Grimani:2002yz} are also plotted.}
\end{fig}

%% file: tex/sect5.tex
\section{Secondary positron flux at Earth}
\label{sec:05}

%
%
\begin{fig}
 \includegraphics[width=0.9\textwidth]{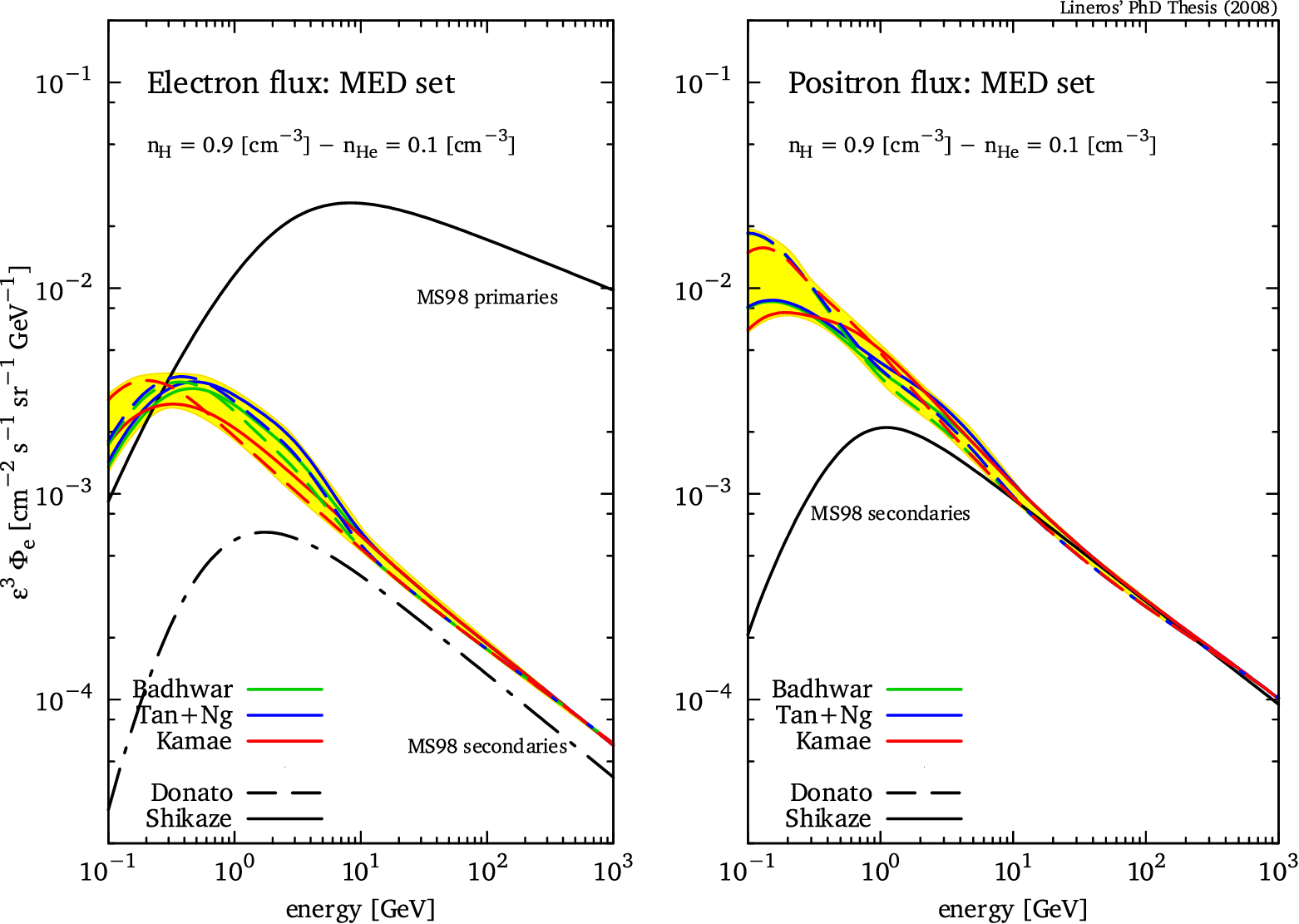}
 \caption{\label{f:ele-pos-flux} Interstellar electron and positron fluxes $\ener{}^{3}\Phi_e$ for the MED set versus energy. The curves correspond to fluxes calculated from Nuclei CR interactions with the ISM. Each curve represents a different nuclear cross section and Nuclei CR flux parameterization. Also the Strong et al.~\cite{Baltz:1998xv} flux parameterizations are shown. The uncertainty band related to those parameterization is plotted (yellow band) as well.}
\end{fig}

The secondary component is the result of the interaction of  cosmic rays nuclei with the interstellar gas composed mainly by hydrogen and helium.
We model the interstellar gas assuming a homogeneous disk with radius equal to the galactic one and thickness of 200~\tu{pc} (details in \cite{Lineros:2008hh}).
%

%
The source term of secondary electrons and positrons coming from interaction of proton cosmic rays with hydrogen is:
\begin{eq}
	q_{p\tn{H}}(\mb{x},\ener{e}) = 4\pi \; n_{\tn{H}}(\mb{x}) \int d\ener{p}\; \Phi_{p}(\mb{x},\ener{p}) \;  \frac{d\ics_{p\tn{H}}}{d\ener{e}} (\ener{p},\ener{e})\;,
\end{eq}
where $n_H$ is the hydrogen number density, $\Phi_{p}$ is the proton flux (details in \cite{Lineros:2008hh}), and $\ics$ represents the inclusive cross section of the process $p + p \rightarrow e^{+}(e^{-}) + X$ discussed and calculated in \citesec{sec:02}.
Let us remained that the inclusive cross section, in our case, comes from the invariant cross section parameterizations of Badhwar et al.~\cite{Badhwar:1977zf}, Tan and Ng~\cite{Tan:1984ha}, and Kamae et al.~\cite{Kamae:2004xx}.\\

Let us stress that the complete positron and electron source term is the sum of all contributions from proton, alpha particles, hydrogen and helium:
\begin{eq}\label{e:sec-srcterm}
	q_{\tn{full}}(\mb{x},\ener{e}) = 4\pi \sum_{i=\tn{H},\tn{He}} \sum_{j=p,\alpha}  \; n_{i}(\mb{x}) \int d\ener{j}\; \Phi_{j}(\ener{j}) \;  \frac{d\ics_{ji}}{d\ener{e}} (\ener{j},\ener{e})\;,
\end{eq}
where the inclusive cross sections for processes $p$+He, $\alpha$+H and $\alpha$+He are estimated by scaling proton--hydrogen cross section:
\begin{eq}
 \frac{d\ics_{ij}}{d\ener{e}} (\ener{j},\ener{e}) =  s_f \; \frac{d\ics_{p\tn{H}}}{d\ener{e}} (\frac{\ener{j}}{A_j},\ener{e}) \; ,
\end{eq}
where $A_j$ is the mass number of the incident particle and $s_f$ are scaling factors~\cite{Orth:1976,Norbury:2006hp}.\\

\subsection{\sc secondary positron and positron fraction}

The propagation of secondary positrons is realized according to the two-zone propagation model (section $\citesec{sec:03}$).
Similar to the case of annihilation of dark matter, we study the effects of uncertainties related to nuclear physics and propagation.
In \citefig{f:ele-pos-flux}, we present secondary electrons and positrons obtained from the three nuclear parameterization: Badhwar et al.~\cite{Badhwar:1977zf}, Tan and Ng~\cite{Tan:1984ha} and Kamae et al.~\cite{Kamae:2004xx}. 
We observe that at high energy $(> 10 \tu{GeV})$ all cases converge because at that energies all three parameterizations converge due to most of experimental data is in this range.
%
At low energy, we observe how differences on the nuclear parameterization affect the propagated positrons and electrons.
Nevertheless, this effect is not so important because it occurs in the energy range where solar modulation dominates.\\ 

%

%
%
\begin{fig}
 \resizebox{0.9\hsize}{!}{\includegraphics[angle=270]{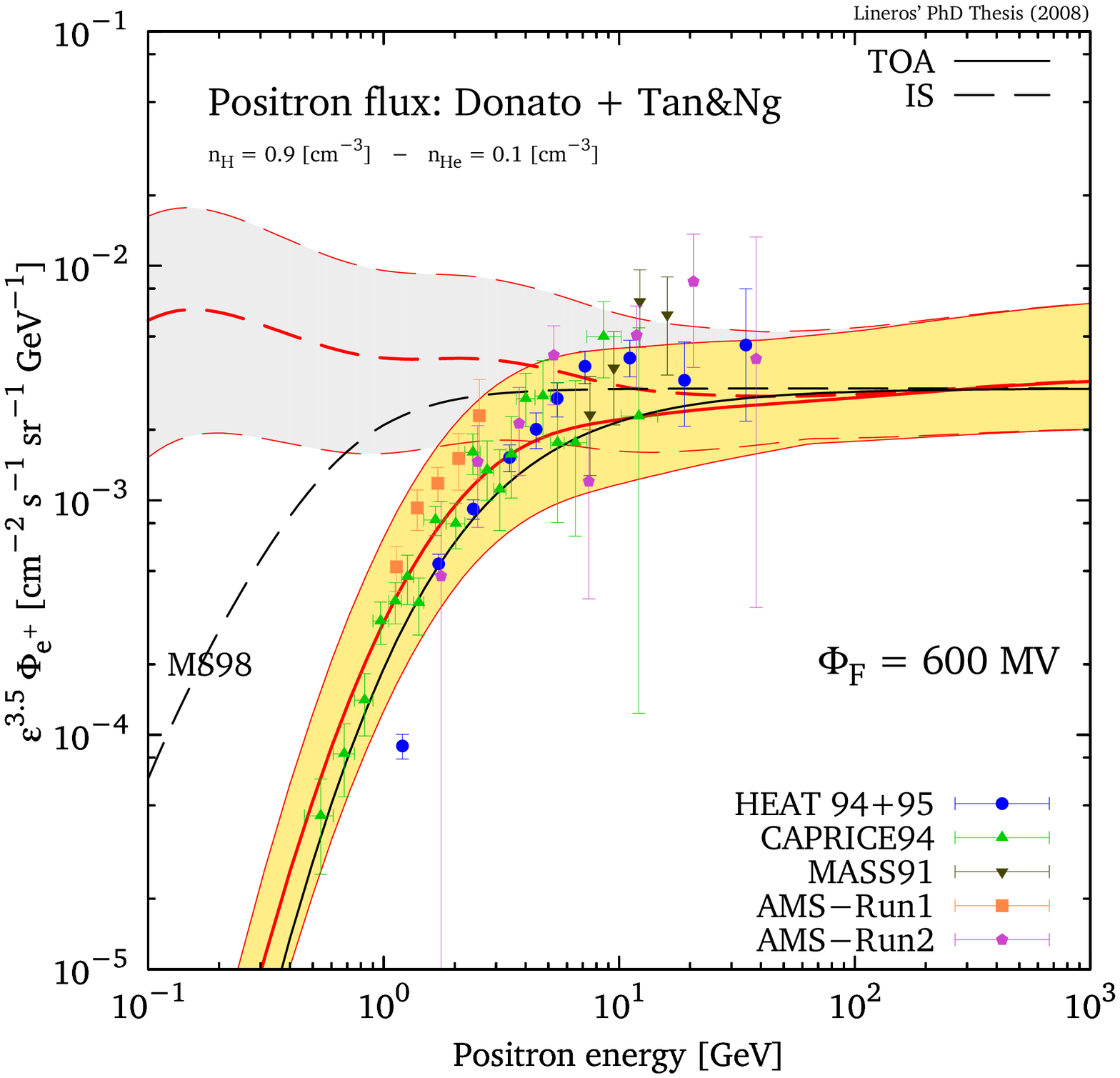}\includegraphics[angle=270]{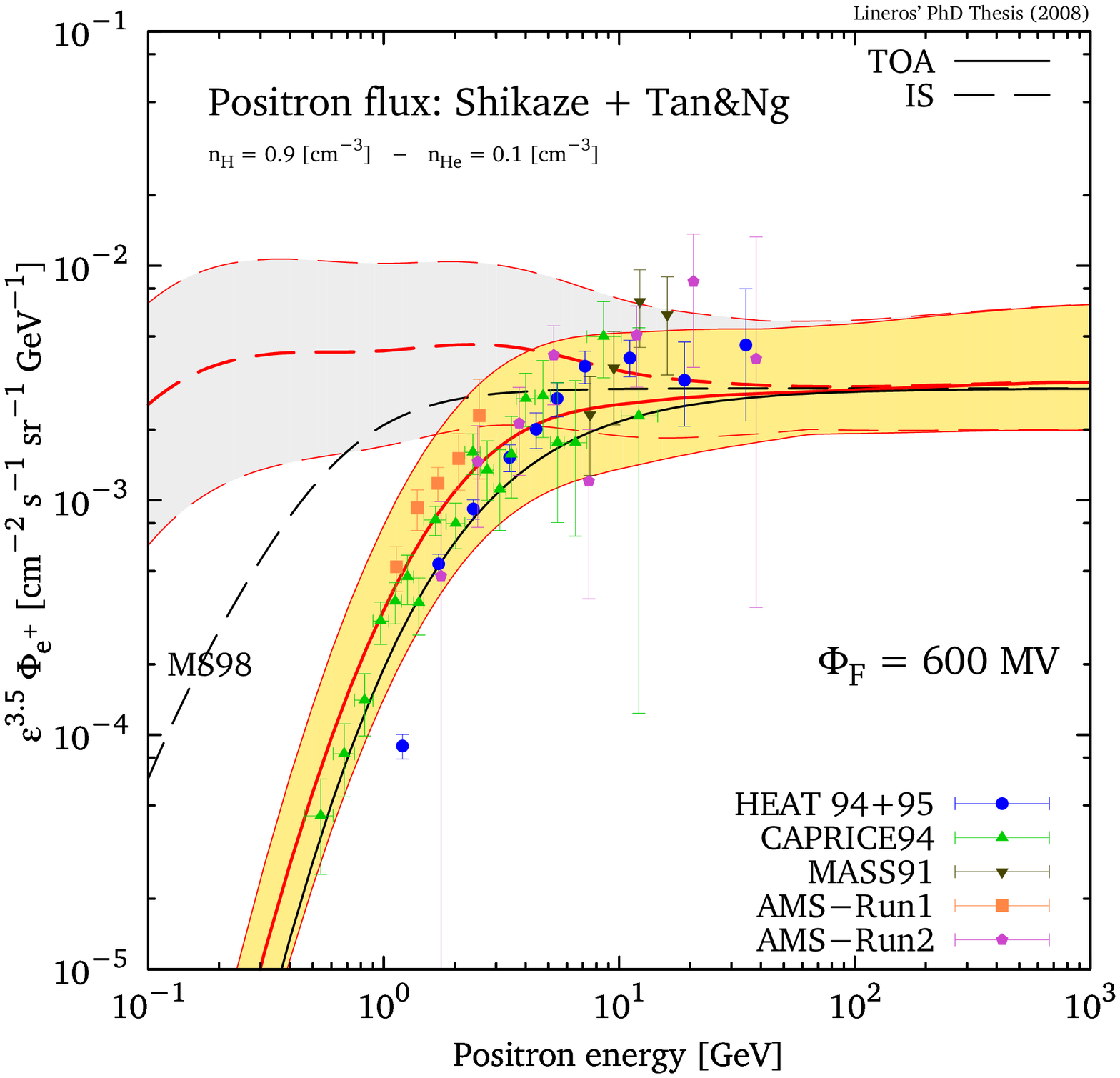}}
 \caption{\label{f:sec-flux-tanng} Secondary positron flux $\ener{}^{3.5} \fluxe$ versus positron energy. Positron fluxes were calculated using proton and alpha CR fluxes from Donato et al.~\cite{Donato:2001ms} and Shikaze et al.~\cite{Shikaze:2006je} with the Tan et al. cross section parameterization~\cite{Tan:1984ha}.}
\end{fig}
%
%
%

%
%
\begin{fig}
\resizebox{0.9\hsize}{!}{\includegraphics[width=\textwidth]{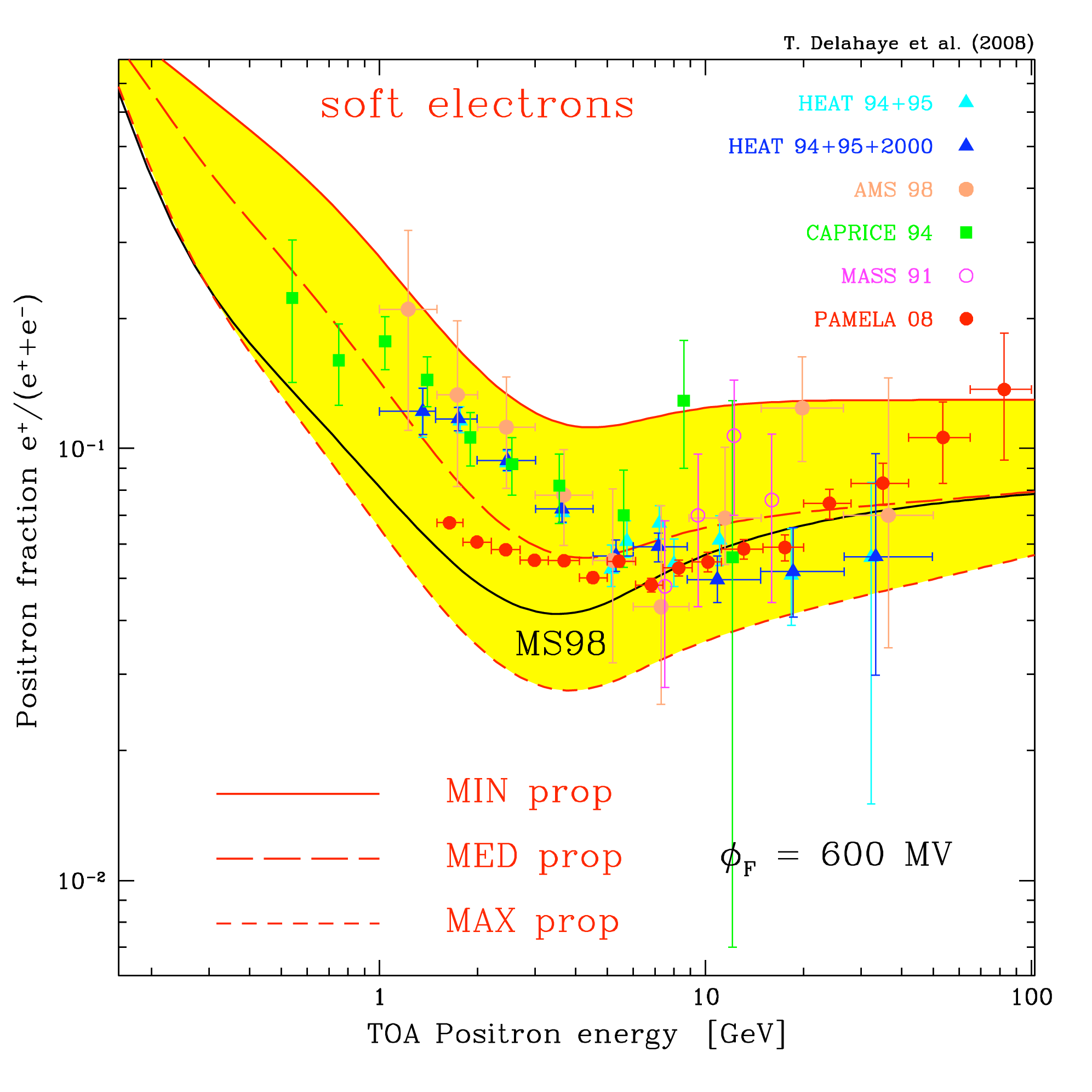}\includegraphics[width=\textwidth]{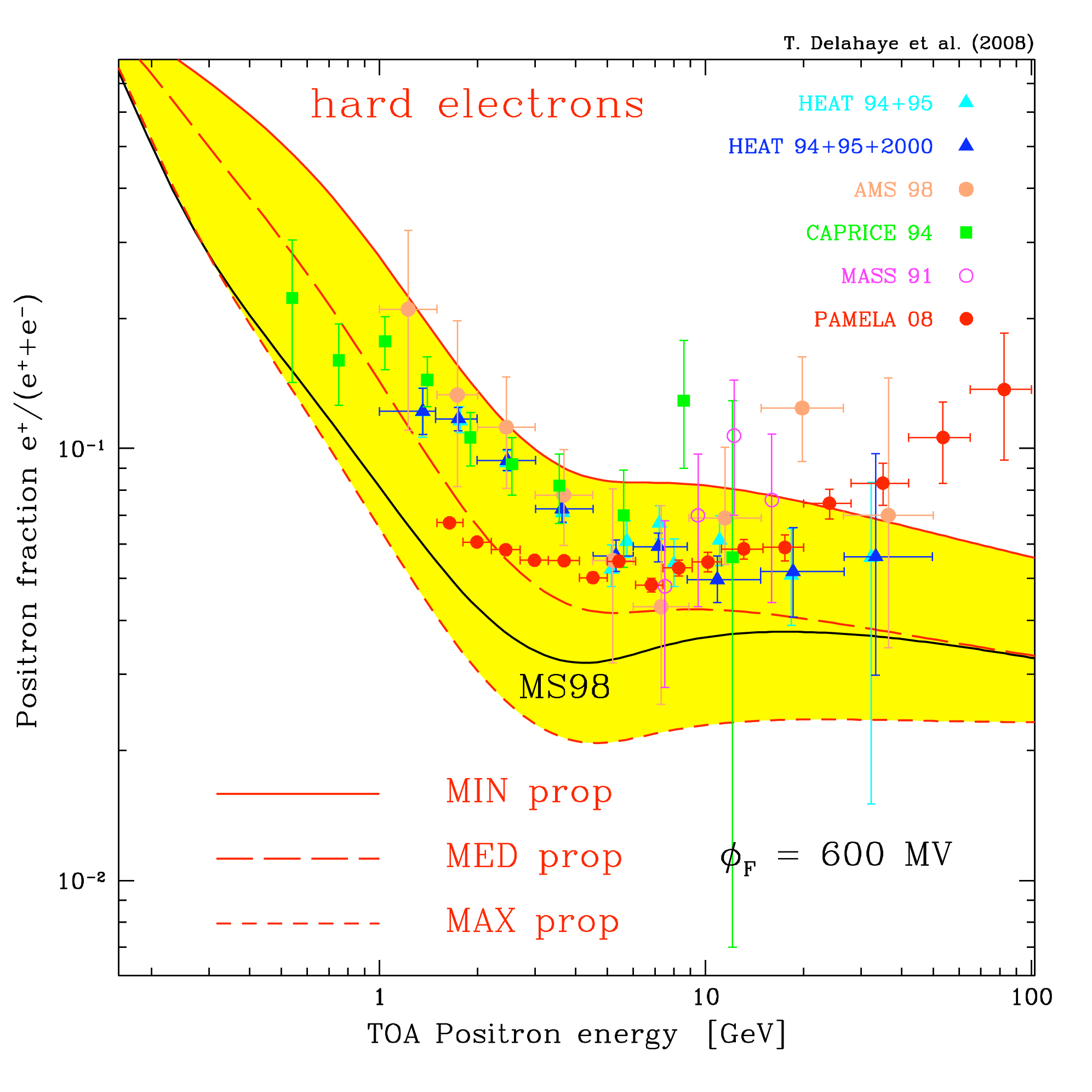}}
	\caption{\label{f:fit-exp-heat} Positron fraction as a function of the positron energy, for a soft and hard electron spectrum. Data are taken from CAPRICE~\cite{Boezio:2000}, HEAT~\cite{Barwick:1997ig}, AMS~\cite{Aguilar:2007}, MASS~\cite{Grimani:2002yz} and PAMELA~\cite{Adriani:2008zr}.}
\end{fig}

Moreover, we can not forget the presence of the uncertainties related to propagation.
\citefig{f:sec-flux-tanng} shows the secondary positron flux versus positron energy.
We consider the whole space of parameter compatible with the B/C analysis~\cite{Maurin:2001sj}, 
 obtaining that the estimated flux encompasses the data. This confirms the compatibility among electron--positron cosmic ray propagation with nuclei cosmic rays case from the propagation point of view.\\
%


The natural extension of the analysis is to calculate the positron fraction. 
Using the calculated secondary flux and the total electron flux estimated from observation like AMS, we observe how the positron fraction is highly sensitive to small variation in the electron flux (\citefig{f:fit-exp-heat}).
We find that harder spectra makes stronger the evidence of positron excess, however a softer electron spectra makes possible to explain the raise in the faction using mainly secondary positrons.\\

This study points towards a more detailed study on the electron flux which has been recently addressed in Ref.~\cite{Delahaye:2010ji}.\\

%% file: tex/sect6.tex
\section{Conclusions}
\label{sec:06}

During the last years, many of the latest cosmic rays experiments have shown very interesting results that are pushing to new frontiers the knowledge about the galactic environment and the origin of the GeV--TeV cosmic rays.
And of course, the case of electron and positron cosmic rays is not an exception.\\

Dark matter annihilation is a very exciting possibility to explain the positron excess, which is not sufficiently explained by contributions from secondary positrons.
In our works, we studied propagation uncertainties associated to the analysis made on cosmic rays nuclei (B/C).
These uncertainties affect considerably the signal coming from annihilation of dark matter, however, most of the characteristics related to the annihilation were not significantly modified.
This fact is promising for further analysis on the signal.\\

To go deeper in the analysis, we studied the secondary production of positrons.
We considered uncertainties related to nuclear physics in addition to the ones related to propagation.
One of the first results was that secondary positron production is compatible with current measurements and with the B/C analysis, which is remarkable considering that we proposed that cosmic ray propagation is common for all the species.\\
In both studies, the positron fraction was analyzed.
Dark matter annihilation scenario is able to explain it, although, the lack of precision from the theoretical point of view makes hard to identify an exotic component present in it.
Moreover, the positron fraction is very sensitive to variations in the electron flux.\\

We stress the necessity to re-estimate secondary and primary electron component, and to consider already known sources, especially pulsars, in order to discard/confirm possible presence of an undiscovered component, like the case of dark matter annihilations.\\

%% file: tex/ack.tex
\acknowledgments
R.L. acknowledges the Comisi\'{o}n Nacional de Investigaci\'{o}n Cient\'{\i}fica y Tecnol\'{o}gica (CONICYT) of Chile for Ph.D. scholarship $\tn{N}^{\circ}$ BECAS-DOC-BIRF-2005-00, the International Doctorate on AstroParticle Physics (IDAPP), and the Istituto Nazionale di Fisica Nucleare (INFN) for the Fubini award.
Work supported by research grants funded jointly by Ministero dell’Istruzione, dell’Universit\'{a} e della Ricerca (MIUR), by Universit\`{a} di Torino (UniTO), by Istituto Nazionale di Fisica Nucleare (INFN) within the Astroparticle Physics Project, by the Italian Space Agency (ASI) under contract $\tn{N}^{\circ}$ I/088/06/0. 

%% file: tex/bib.tex
\bibliographystyle{h-elsevier3}
\bibliography{bibliography} 